\documentclass[aps,prb,twocolumn,superscriptaddress,floatfix,10pt]{revtex4-1}

\usepackage{graphicx}
\usepackage{amsmath}
\usepackage{amssymb}
\usepackage{xcolor}
\usepackage[bookmarksnumbered=true,hypertexnames=false,colorlinks=true,linkcolor=blue,urlcolor=blue,citecolor=blue,anchorcolor=green]{hyperref}
\usepackage[normalem]{ulem}

\newcommand{\bra}[1]{\mbox{$\langle #1 |$}}
\newcommand{\ket}[1]{\mbox{$| #1 \rangle$}}
\newcommand{\braket}[2]{\mbox{$\langle #1 | #2 \rangle$}}

\newcommand{\norm}[1]{\mbox{$\left\| #1 \right\|$}}
\newcommand{\ovr}[1]{\mbox{$\overset{\rightharpoonup}{#1}$}}
\newcommand{\ovl}[1]{\mbox{$\overset{\leftharpoonup}{#1}$}}

\begin{document}

\title{Partitioned Expansions for Approximate Tensor Network Contractions}
\author{Glen Evenbly}
\email{evenbly@amazon.com}
\affiliation{AWS Center for Quantum Computing, Pasadena, CA 91125, USA}
\author{Johnnie Gray}
\author{Garnet Kin-Lic Chan}
\affiliation{Division of Chemistry and Chemical Engineering,  California Institute of Technology, Pasadena, CA 91125, USA}
\date{\today}

\begin{abstract}
We propose a method for approximating the contraction of a tensor network by partitioning the network into a sum of computationally cheaper networks. This method, which we call a \emph{partitioned network expansion} (PNE), builds upon recent work that systematically improves belief propagation (BP) approximations using loop corrections~\cite{evenbly2025loopseriesexpansionstensor}. However, in contrast to previous approaches, our expansion does not require a known BP fixed point to be implemented and can still yield accurate results even in cases where BP fails entirely. The flexibility of our approach is demonstrated through applications to a variety of example networks, including finite $2D$ and $3D$ networks, infinite networks, networks with open indices, and networks with degenerate BP fixed points. Benchmark numerical results for networks composed of Ising, AKLT, and random tensors typically show an improvement in accuracy over BP by several orders of magnitude (when BP solutions are obtainable) and also demonstrate improved performance over traditional network approximations based on singular value decomposition (SVD) for certain tasks.
\end{abstract}

\maketitle
\tableofcontents


\section{Introduction} \label{sect:introduction}
Tensor networks\cite{ORUS2014117,Bridgeman2017Waving,Ran2020Contractions} are powerful tools for the classical simulation of quantum many-body systems and, in many cases, provide the most effective known classical simulation algorithms. More recently, they have also been used to benchmark nascent quantum computers\cite{Fried2018Handler,Villalonga2019Verify,Schutski2020Adaptive,Zhou2020Limits,levental2021FPGA,Vincent2022Jet,Pan2022BigBatch,Pan2022Sycamore,Ayral2023Fidelity,Kim2023Evidence,Begusic2024,Tindall2024ibm,Orus2024ibm,tindall2025dynamicsdisorderedquantumsystems}, where claims of quantum supremacy are often tested against state-of-the-art tensor network methods. A key component of any tensor network algorithm is the process of network contraction: it is required both for extracting information from a network and for optimizing its tensors according to some desired criteria (e.g., minimizing the energy of a quantum state represented by the network). Often approximations are used to reduce the computational cost of tensor network contractions, given that exact contractions might otherwise be untenable, with approximations based on the singular value decomposition (SVD) being the most commonplace\cite{Kolda2009,gray2021hyper}.

In the last few years, ideas from belief propagation (BP)\cite{Bethe1935Stats,Pearl2022Reverend,Pearl1986Fusion}, a message passing algorithm with a long history in both computer science\cite{Mezard2002Satis,Yedidia2003BP,Yedidia2005Free} and in statistical physics\cite{mezard1987spin}, have emerged as a promising alternative for approximating tensor network evaluations\cite{Leifer2008Graphical,Poulin2008BP,Wrigley2017TBP,AlkabetzArad2021,Sahu2022Sparse, Tindall2023Scipost,Guo2023BlockBP,Wang2023TNMP,pancotti2023one}. The basic approach works by iterating BP message passing until converged to a fixed point, after which the fixed-point messages are inserted into the network to obtain an approximation in which the bond dimensions have been effectively reduced to trivial rank ($r=1$). These BP-based approximations offer several appealing features: they are exceptionally computationally efficient and can be applied directly to networks that are difficult to handle using conventional SVD-based approximations, such as $3D$ networks and unstructured networks. However, although often qualitatively accurate, the BP approximation is insufficiently precise for many practical applications. To address this limitation, strategies have been developed to improve accuracy by constructing an expansion around the BP fixed point\cite{evenbly2025loopseriesexpansionstensor,park2025simulatingquantumdynamicstwodimensional,midha2025beliefpropagationclustercorrectedtensor,gray2025tensornetworkloopcluster}, with the aim of incorporating the dominant corrections that BP neglects. These expansion methods, which in principle can achieve arbitrarily high precision through the inclusion of higher-order terms, have already proven useful in the classical simulation of quantum experiments that appear intractable with previous techniques\cite{tindall2025dynamicsdisorderedquantumsystems}.

Network contraction methods based on expansions around a BP fixed point still face significant limitations: (i) they require a known BP fixed point as a starting point and therefore cannot be applied when message passing fails to converge, and (ii) even when a BP fixed point is obtained, they may still fail for networks that possess multiple fixed points of comparable weight. In this work, we formulate a more general series expansion for tensor networks, which we call a partitioned network expansion (PNE). Although it shares some features with previous approaches, the PNE can circumvent limitations (i) and (ii). In particular, the PNE can be applied to networks without a known BP fixed point, provided that one has some knowledge of the dominant subspace associated with the network indices. The proposed expansion can also employ higher-rank projections on each index (whereas the subspace formed from BP messages is inherently rank $r=1$), enabling accurate results even for networks with competing BP fixed points. Additionally, the PNE is flexible, allowing the level of approximation to be tailored to a given computational budget, and is relatively straightforward to implement numerically.

The paper is organized as follows. In Sect.~\ref{sect:partition} we describe the general formulation of partitioned expansions and discuss the various ways in which they can be applied. In Sect.~\ref{sect:small} we provide examples of how the expansion may be applied to small tensor networks, analyze the resulting error terms, and present benchmark numerical results demonstrating the efficacy of the method. More sophisticated uses of the PNE are explored in Sect.~\ref{sect:large}, including examples of recursive applications of the expansion and comparisons against SVD truncation. We also describe its use for evaluating networks of infinite extent. In Sect.~\ref{sect:beyond} we examine expansions based on higher-rank projections, where the expansions are not tied to BP fixed points, and demonstrate successful application of the PNE to a network in which BP-based methods fail due to a degeneracy of BP fixed points. Finally, key aspects of the method and future directions are discussed in Sect.~\ref{sect:discussion}.

\section{Partitioned Expansions} \label{sect:partition}
\subsection{General Formulation} \label{sect:formulation}
\begin{figure} [!t] 
\begin{center}
\includegraphics[width=8.0cm]{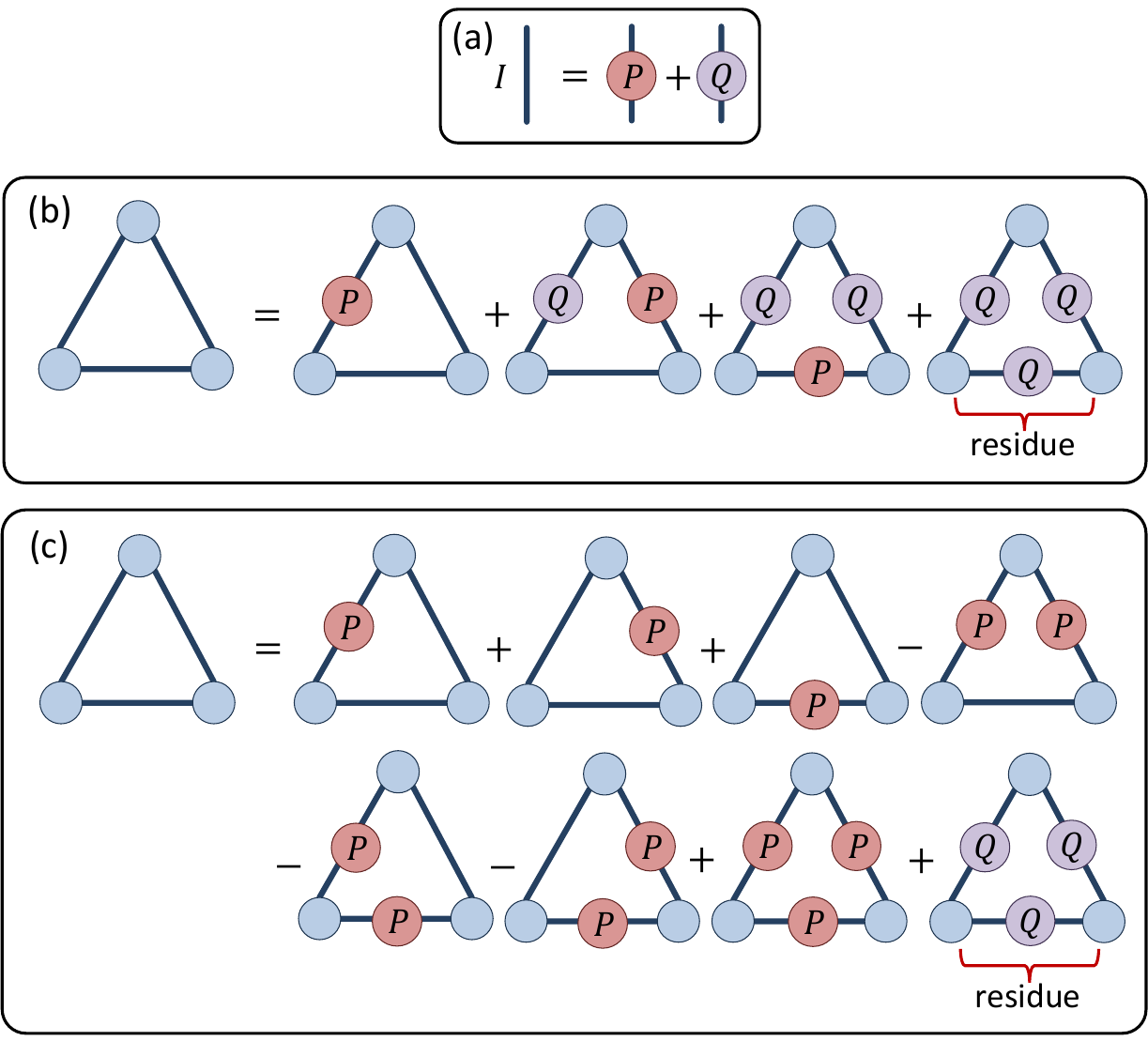}
\caption{(a) The identity $I$ is decomposed as the sum of a projector $P$ and its compliment $Q$. (b) The the linear form of the partitioned network expansion for a network of three tensors, see also Eq.~\ref{eq:A3}. (c) The combinatorial form of the expansion, see also Eq.~\ref{eq:A5}}
\label{fig:1}
\end{center}
\end{figure}

In this section, we describe a general strategy for approximating the contraction of a tensor network as a sum over a set of computationally cheaper network contractions, using a method that we call a \emph{partitioned network expansion} (PNE). Let $\mathcal T$ be a closed tensor network consisting of tensors $\{T_1, T_2, T_3, \ldots, T_N\}$ connected by indices labeled $\{i, j, k, \ldots\}$. Suppose our goal is to contract $\mathcal T$ to a scalar $\mathcal{Z}_0$,
\begin{equation}
\mathcal{Z}_0 = \sum_{i,j,k,\ldots} T_1 (\xi_1) \ T_2 (\xi_2) \ T_3 (\xi_3) \ldots, \label{eq:A1}
\end{equation}
where $\xi_r$ denotes the subset of indices $\xi_r \subseteq \{i, j, k, \ldots\}$ connected to the tensor $T_r$. We now consider a modified network formed by inserting operators $\hat o_i, \hat o_j, \hat o_k, \ldots$ onto the edges $i, j, k, \ldots$ of the original network, and define $\mathcal Z(\hat o_i, \hat o_j, \hat o_k, \ldots)$ as the scalar resulting from contracting the modified network. Let $P_i$ be a projector acting on index $i$ and $Q_i$ its complement, such that $Q_i + P_i = I_i$, where $I_i$ is the identity operator of dimension $\norm{i}$. It follows that the original network may be decomposed into two contributions, one with projector $P_i$ on index $i$ and one with projector $Q_i$:
\begin{equation}
\mathcal Z_0 = \mathcal Z(P_i, I_j, I_k) + \mathcal Z(Q_i, I_j, I_k, \ldots). \label{eq:A2}
\end{equation}
In this context, the projectors $P_i$ and $Q_i$ effect a bipartition of index $i$. For reasons that will become clear later, we refer to the subspaces captured by $P_i$ and $Q_i$ as the dominant and subordinate subspaces, respectively. This splitting into $P$ and $Q$ contributions can be iterated over other indices, using different projectors on each index if desired. For a network with three indices $\{i, j, k\}$ we obtain
\begin{align}
\mathcal Z_0 &= \mathcal Z(P, I, I) + \mathcal Z(Q, P, I)  \nonumber\\
& + \mathcal Z(Q, Q, P) + \mathcal Z(Q, Q, Q), \label{eq:A3}
\end{align}
where the operator subscripts have been omitted for clarity, see also Fig.~\ref{fig:1}(b). More generally, for a network with $M$ indices, the (linear form of the) partitioned network expansion can be written as
\begin{equation}
\mathcal Z_0 = \mathcal Z\left(Q^{\otimes M}\right) + \sum_{r = 0}^{M-1} \mathcal Z\left(Q^{\otimes r} \otimes P \right). \label{eq:A4}
\end{equation}
Here the first term denotes the case in which $Q$ is inserted on all $M$ indices of the network, which we call the \emph{residue} $\mathcal R$,
\begin{equation}
\mathcal R \equiv \mathcal Z \left(Q^{\otimes M}\right), \label{eq:res}
\end{equation}
which can equivalently be interpreted as projecting every index of the network into the subordinate subspace. Similarly, each term $\mathcal Z \left(Q^{\otimes r} \otimes P\right)$ from Eq.~\ref{eq:A4} represents the network obtained by inserting $r$ copies of $Q$ followed by one insertion of $P$ (and implicitly followed by the identity on the remaining indices).

In some cases it is convenient to work with a \emph{combinatorial form} of the partitioned network expansion, which is obtained by further expanding the sub-networks (excluding the residue) so that they contain only $P$ projector terms. For the three-index network in Eq.~\ref{eq:A3}, this yields
\begin{align}
\mathcal Z_0 &= \mathcal Z(P, I, I) + \mathcal Z(I, P, I) + \mathcal Z(I, I, P) \nonumber\\
& - \mathcal Z(P, P, I) - \mathcal Z(I, P, P) - \mathcal Z(P, I, P) \nonumber\\
&+ \mathcal Z(P, P, P) + \mathcal Z(Q, Q, Q) \label{eq:A5}
\end{align}
see also Fig.~\ref{fig:1}(c). More generally, for a network with $M$ indices, the combinatorial PNE expands the network as a sum over $2^M$ sub-networks,
\begin{equation}
\mathcal Z_0 = \mathcal Z\left(Q^{\otimes M}\right) + \sum_{|x| \textrm{\ odd}} \mathcal Z (P^{\otimes x})  - \sum_{\substack{|x|\ \text{even}\\ |x|>0}} \mathcal Z(P^{\otimes x}). \label{eq:A6}
\end{equation}
Here $x$ denotes the set of $M$-length binary strings $x\in \{0,1\}^M$, such that $P^{\otimes x}$ is a binary string composed of $P$'s and $I$'s. Thus Eq.~\ref{eq:A6} represents, in addition to the usual residue $\mathcal R$ of Eq.~\ref{eq:res}, the sum over all configurations with an odd number of $P$ insertions into the network minus the sum over all configurations with an even number of $P$ insertions (excluding the case with zero insertions). Finally, we remark that it is possible to apply the expansion only to a subset of the indices in the lattice rather than to all indices. Consequently, the variable $M$ in Eqs.~\ref{eq:A4} and \ref{eq:A6} should, in general, be interpreted as the total number of partitions being used, rather than the total number of indices in the network.

\subsection{Construction of Projectors} \label{sect:projectors}
The two forms of the PNE given in Eq.~\ref{eq:A4} and Eq.~\ref{eq:A6} represent different ways of expanding a single network as a sum of networks by partitioning indices into orthogonal components (i.e., components described by complementary $P$ and $Q$ projectors). We emphasize that these equations constitute exact identities, valid for any network $\mathcal T$ and any choice of complementary $P$ and $Q$ projectors. In practice, however, we use these expansions to approximate tensor network contractions by neglecting the residue term $\mathcal R$ of Eq.~\ref{eq:res}. For the PNE to be computationally useful, two basic requirements should be satisfied:
\begin{enumerate}
    \item The projectors $P$ and $Q$ should be chosen so that the neglected residue term $\mathcal R$ is small, ensuring that the contraction $\mathcal Z_0$ is accurately approximated.
    \item The projectors $P$ should be chosen to have low rank, so that evaluating the sum of terms in the expansion is computationally cheaper than evaluating the original network.
\end{enumerate}
It is not immediately obvious under what conditions these requirements can be met for a generic tensor network. As a general guiding principle, one may attempt to satisfy the first condition by choosing each projector $P$ to project onto the \emph{dominant subspace} of an index, i.e., the subspace that contributes the largest-magnitude terms to the sum $\mathcal Z_0$ in Eq.~\ref{eq:A1}.

In many cases, suitable rank $r=1$ projectors $P$ can be constructed from fixed-point BP messages, which we now discuss in more detail. Let us assume that a BP fixed point has been obtained for the network under consideration, so that for each index $i$ we possess an incoming message $\ket{\ovr{\mu_i}}$ and an outgoing message $\ket{\ovl{\mu_i}}$ (relative to some predefined orientation of the index). In general, the incoming and outgoing messages on each index will not be equal. However, as described in Appendix~\ref{sect:sym}, it is always possible to re-gauge each index $i$ in such a way that these messages are symmetrized, i.e., both the incoming and outgoing messages become equal to some new message $\ket{\mu_i}$. A rank $r=1$ projector $P_i$ can then be constructed from the outer product of this symmetrized message,
\begin{equation}
P_i = \ket{\mu_i}\bra{\mu_i}. \label{eq:A7}
\end{equation}
When constructing the projectors $P$ from fixed-point BP messages, the PNE can be interpreted as being closely related to recent proposals for improving the accuracy of BP by incorporating loop corrections\cite{evenbly2025loopseriesexpansionstensor,park2025simulatingquantumdynamicstwodimensional,midha2025beliefpropagationclustercorrectedtensor,gray2025tensornetworkloopcluster}, though it still retains several significant advantages, as discussed in Sect.~\ref{sect:potential}.

Importantly, unlike these other proposals for improving BP accuracy, the present approach does not intrinsically require a known BP fixed point, since the PNE is valid for any choice of $P$. Methods for determining projectors $P$ that do not rely on first obtaining a BP fixed point are discussed in Sect.~\ref{sect:weight}. Thus, we can still apply the PNE to systems for which no BP fixed point is known (including systems that fail to converge to a fixed point under iterations of the message passing algorithm), whereas previous BP-based approaches are simply inapplicable in such settings. Our method also permits the use of projectors $P$ with rank $r>1$, providing an additional means to tune the accuracy of the approximation. This flexibility is especially important for networks that possess multiple competing BP fixed points of equal (or nearly equal) weight where, even when a BP fixed point can be found, the expansion about a single BP fixed point will generally be insufficient. An example of this behavior appears in the subcritical Ising model, examined in Sect.~\ref{sect:degen}, which exhibits two BP fixed points of equal weight, such that an expansion using rank $r\ge 2$ projectors is required to obtain accurate results.

\subsection{Multi-index Partitions} \label{sect:multi}
For reasons of simplicity, the PNE was formulated in Sect.~\ref{sect:formulation} in terms of projectors $P_i$ acting on individual indices $i$ of a network. In practice, however, it is often useful to extend this formalism to projectors that act on multiple indices. For example, one may employ a pair of complementary projectors $P_{ij}$ and $Q_{ij}$ that act on the pair of indices $\{i, j\}$; in general, we refer to this as a \emph{multi-index partition}. Some care is required when using multi-index partitions, since inserting a multi-index projector can increase the contraction cost scaling of the network (whereas insertion of single-index projectors never increases the leading-order cost). To avoid such increases in cost scaling, we will often use multi-index projectors $P_{ij}$ that factorize as a product of single-index projectors,
\begin{equation}
P_{ij} = P_i \otimes P_j. \label{eq:A8}
\end{equation}
Note that a multi-index partition with a factorized projector of the form Eq.~\ref{eq:A8} is not equivalent to introducing two single-index partitions using $P_i$ and $P_j$ separately. In the former case, $P_i$ and $P_j$ always appear together within the expansion terms. Moreover, the complement $Q_{ij}$ of $P_{ij}$ does not factorize into a product of single-site components
\begin{equation}
Q_{ij} = (Q_i \otimes P_j) + (P_i \otimes Q_j) + (Q_i \otimes Q_j). \label{eq:A9}
\end{equation}
Consequently, expansion terms involving insertions of $Q_{ij}$ may still lead to increased contraction cost. For this reason, when employing multi-index partitions we utilize the combinatorial form of the PNE defined in Eq.~\ref{eq:A6}, since this form avoids the explicit use of complementary $Q$ projectors. The form of the residue $\mathcal R$ from the PNE, previously defined in Eq.~\ref{eq:res}, also becomes more complicated when using multi-index partitions. For a multi-index partition involving indices $\{i, j\}$ and a factorized projector $(P_i \otimes P_j)$, the residue includes contributions in which $i$ and $j$ are projected individually into the subordinate subspace, as follows directly from Eq.~\ref{eq:A9}. When employing multi-index partitions that factorize as in Eq.~\ref{eq:A8}, it is often useful to construct the single-site projectors from fixed-point BP messages $\ket{\mu}$, such that
\begin{equation}
P_{ij} = \ket{\mu_i}\bra{\mu_i} \otimes \ket{\mu_j}\bra{\mu_j}. \label{eq:A10}
\end{equation}
In this setting, a partition can be represented as a line that transects one or more indices of a tensor network, where each transected index is cut and then capped by the appropriate BP message. Examples of this representation are shown in Fig.~\ref{fig:4}.

\subsection{Potential Advantages} \label{sect:potential}
Below we summarize several advantages of the PNE that will be demonstrated throughout this manuscript, both in comparison with previous BP-based tensor network approximations and with truncation-based approaches:
\begin{enumerate}
    \item The method does not require a known BP fixed point in order to be applied.
    \item By utilizing higher-rank projectors, rank $r>1$, the approach can approximate networks containing multiple competing BP fixed points in situations where expansions based on a single BP fixed point would otherwise fail.
    \item The method offers a high degree of flexibility in choosing which indices to partition. In practice, this allows users to implement an approximation consistent with their computational budget. For example, for the network shown in Fig.~\ref{fig:6}, which has an exact contraction cost of $O(\chi^6)$, one can construct different approximations that scale as $O(\chi^5)$, $O(\chi^4)$, or $O(\chi^3)$.
    \item The proposed expansion is not restricted to tensor networks that evaluate to a scalar; it can equally be applied to networks with open indices, such as those yielding transfer matrices or density matrices, which are particularly useful in tensor network optimization algorithms.
    \item Partitioned expansions are relatively straightforward to automate in code, especially compared with previous proposals that require enumerating and evaluating loop corrections individually.
    \item All terms in the expansion can be evaluated in parallel, allowing the method to exploit modern computational resources more effectively. In contrast, truncation-based tensor network approximations (i.e., those based on SVD\cite{gray2021hyper}) are typically implemented as a serial sequence of alternating contractions and decompositions, which is difficult to parallelize.
\end{enumerate}
Much of the remainder of this manuscript focuses on providing and analyzing examples that demonstrate the points outlined above.

\begin{figure}[!t] 
\begin{center}
\includegraphics[width=8.5cm]{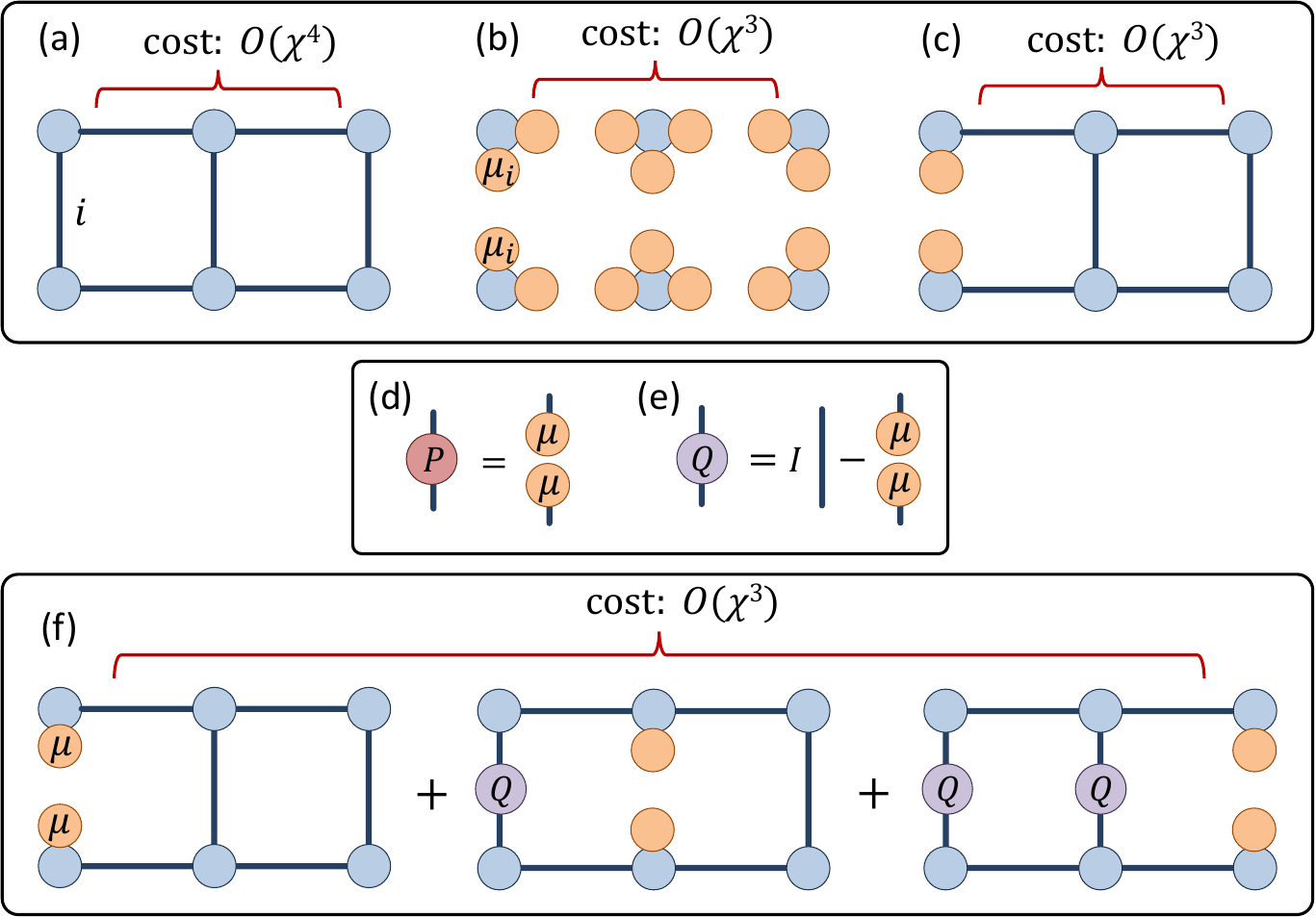}
\caption{(a) A network of six tensors, where indices are assumed to be $\chi$-dimensional, has an exact contraction cost of $O(\chi^4)$. (b) The BP approximation, whose contraction cost scales as $O(\chi^3)$, is obtained by inserting fixed-point message pairs $\ket{\mu_i}$ on all indices $i$. (c) A network with contraction cost $O(\chi^3)$ can also be obtained by inserting the BP messages on a single index. (d) Rank $r=1$ projectors $P$ can be formed from the outer product of the fixed-point BP messages. (e) Complementary projectors $Q$ can be formed by subtracting $P$ from the identity $I$. (f) An $O(\chi^3)$ approximation is obtained via the linear form of a PNE based on partitioning the three vertical indices.}
\label{fig:2}
\end{center}
\end{figure}

\section{Small Network Examples} \label{sect:small}
\subsection{Double-loop Network} 
As a first example we consider a network composed of six tensors as depicted in Fig.~\ref{fig:2}(a), with bond dimension $\chi$ on all indices. It is easily determined that this network can be exactly contracted to a scalar $\mathcal Z_0$ with a cost that scales as $O(\chi^4)$. Let us assume that our goal is now to approximately contract the network at reduced computational cost. 

Suppose that BP message passing has been performed on the network, with cost scaling as $O(\chi^3)$, and that it converges to a BP fixed point. The resulting fixed-point messages can then be symmetrized as described in Sect.~\ref{sect:sym}, so that for each index $i$ in the network we obtain a normalized message $\ket{\mu_i}$ satisfying $\braket{\mu_i}{\mu_i} = 1$. The standard BP approximation to $\mathcal Z_0$, with cost scaling $O(\chi^3)$, is obtained by inserting these fixed-point messages on every index, as shown in Fig.~\ref{fig:2}(b). However, in this particular instance, an $O(\chi^3)$ approximation can also be achieved by inserting the fixed-point messages on only a single index, as illustrated in Fig.~\ref{fig:2}(c). Alternatively, the network may be approximated via a PNE by constructing projectors $P$ and $Q$ from the fixed-point messages as in Eq.~\ref{eq:A7}; see also Fig.~\ref{fig:2}(d--f). We choose to partition only the three vertical indices for reasons that will become evident when we later analyze the residue $\mathcal R$. The resulting expansion contains three networks, each with contraction cost $O(\chi^3)$.

\begin{figure}[!t] 
\begin{center}
\includegraphics[width=8.5cm]{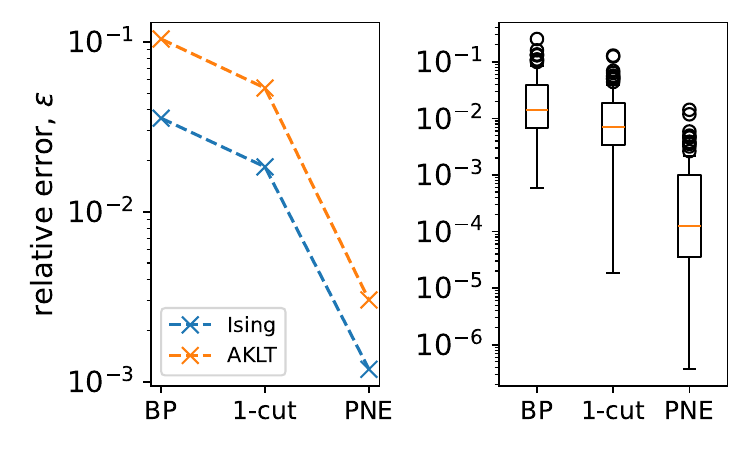}
\caption{Relative errors $\varepsilon$, as defined in Eq.~\ref{eq:B1}, for the contraction of the network shown in Fig.~\ref{fig:2}(a). Results are compared across the BP approximation from Fig.~\ref{fig:2}(b), the single-index cut from Fig.~\ref{fig:2}(c), and the PNE approximation from Fig.~\ref{fig:2}(f). (left) Results obtained using tensors derived from either the $2D$ classical Ising model (at criticality) or the $2D$ square-lattice AKLT model, both blocked into networks of bond dimension $\chi=64$. (right) Aggregate results over 100 independent instances of random tensors with bond dimension $\chi=64$.}
\label{fig:2b}
\end{center}
\end{figure}

We now perform benchmark calculations to compare the accuracy of these different approximation strategies. As described in Appendix~\ref{sect:construction}, we generate the tensors used in the test networks using three different approaches: (i) from the $2D$ classical Ising model at criticality, (ii) from the ground-state projected entangled pair states (PEPS)\cite{Nishino1996Corner,verstraete2004renormalization,Orus2009Corner} representation of the $2D$ square-lattice AKLT model\cite{Affleck1987AKLT,Wei2015AKLT2D}, and (iii) by populating tensors with random elements. In all cases we use tensors with bond dimension $\chi=64$, which—in cases (i) and (ii)—are formed by appropriately grouping patches of tensors from the original $\chi=2$ and $\chi=4$ networks. In Fig.~\ref{fig:2b} we compare the relative error $\varepsilon$ in the scalar network contraction,
\begin{equation}
\varepsilon = \left| \frac{\mathcal Z_\textrm{ex.} - \mathcal Z_\textrm{approx}}{\mathcal Z_\textrm{ex.}} \right|, \label{eq:B1}
\end{equation}
for the various approximations. These results show that cutting a single index, as in Fig.~\ref{fig:2}(c), yields only a modest improvement over the basic BP approximation. In contrast, the PNE produces a substantially larger improvement in accuracy across all test cases.

\begin{figure}[!t] 
\begin{center}
\includegraphics[width=8.5cm]{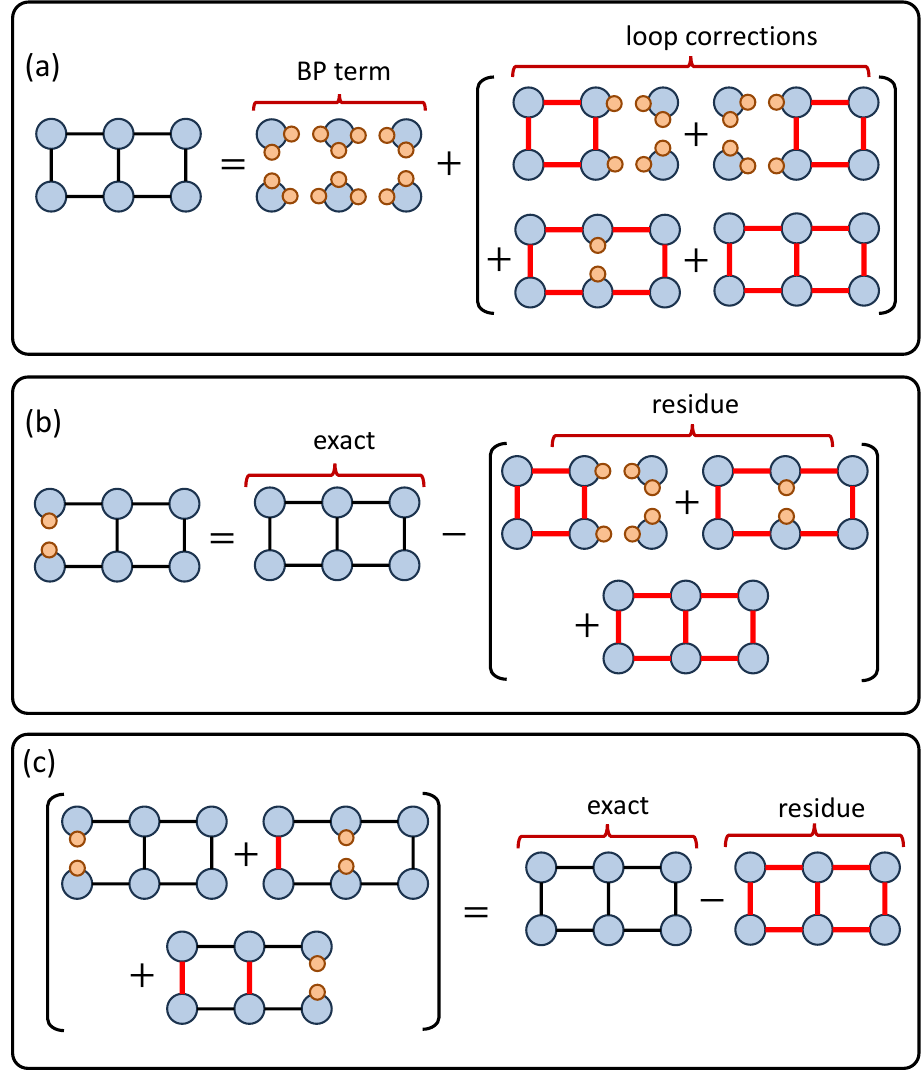}
\caption{(a) Expansion of the network into the BP fixed point plus loop-correction terms, where thick red lines denote indices projected into the subspace orthogonal to the fixed-point messages. (b) Depiction of the residue arising from the approximation in Fig.~\ref{fig:2}(c), which includes contributions of degrees $\{4,6,7\}$. (c) The residue from the PNE in Fig.~\ref{fig:2}(f) contains only a single degree-7 term.}
\label{fig:3}
\end{center}
\end{figure}

The relative magnitudes of the observed errors can be understood in terms of the loop corrections to the BP fixed point\cite{Chertkov2006LoopCalculus,Chertkov2006LoopSeries,evenbly2025loopseriesexpansionstensor}. Using the terminology of Ref.~\onlinecite{evenbly2025loopseriesexpansionstensor}, if an exact BP fixed point is known for a tensor network, then the network can be expanded as the BP vacuum plus loop-correction terms. The loop corrections correspond to configurations in which one or more indices are projected into the \emph{excited} subspace (i.e., the subspace orthogonal to the BP fixed-point messages). It is known that only the loop configurations without dangling excitations contribute nontrivially to the network contraction, where a dangling excitation is defined as a tensor possessing exactly $n=1$ excited index. This decomposition of the network from Fig.~\ref{fig:2}(a) into the BP vacuum plus nontrivial loop corrections is illustrated in Fig.~\ref{fig:3}(a).

In such expansions it is typically observed that the lowest-degree loop configurations, i.e., those with the fewest excited indices, typically provide the dominant corrections beyond the BP vacuum. In the expansion shown in Fig.~\ref{fig:3}(a) the two degree-4 configurations, each corresponding to a loop of excitations around a four-tensor plaquette, are therefore expected to dominate. However, in the single-cut approximation of Fig.~\ref{fig:2}(c), one of these degree-4 configurations is truncated (along with all higher-degree terms), as illustrated in Fig.~\ref{fig:3}(b). This explains why the single-cut approximation shows only a modest improvement over BP in the numerical benchmarks: one of the leading correction terms is missing. In contrast, the PNE truncates only the degree-7 correction term, as shown in Fig.~\ref{fig:3}(c), and therefore exhibits a significantly larger improvement in accuracy. Including additional partitioned indices in the expansion from Fig.~\ref{fig:2}(f) would not improve accuracy, since the nontrivial contribution to the residue would remain unchanged. This is why only three indices were chosen for partitioning.

Although the argument above is formally valid only when the projectors $P$ are constructed from fixed-point BP messages, it nonetheless provides a useful guiding principle for selecting which indices to partition when implementing the PNE. Because the residue $\mathcal R$ represents the sum over configurations in which all partitioned indices are simultaneously projected into the subordinate (or excited) subspace defined by the $Q$ projectors, one should partition indices that are widely separated within the network, thereby maximizing the degree of any correction term appearing in the residue.

\begin{figure}[!t] 
\begin{center}
\includegraphics[width=8.5cm]{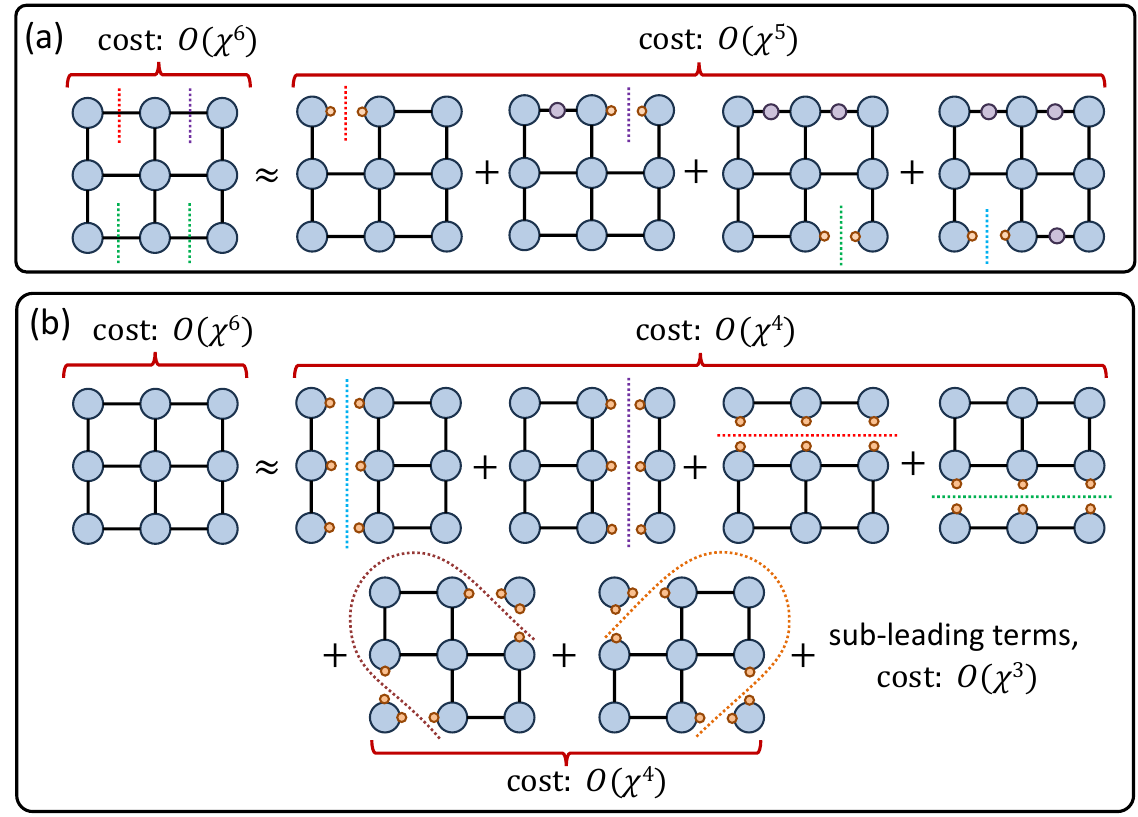}
\caption{(a) The $3\times 3$ network, which has an exact contraction cost of $O(\chi^6)$, is approximated as a sum of cost $O(\chi^5)$ networks via a PNE using four single-index partitions in total. (b) The network is approximated as a sum of cost $O(\chi^4)$ networks via the combinatorial form of a PNE using six multi-index partitions. Sub-leading terms arising from combinations of these partitions are not shown.}
\label{fig:4}
\end{center}
\end{figure}

\subsection{$3 \times 3$ Network} \label{sect:three}
It is instructive to consider another example involving a small tensor network, this time the $3\times 3$ network shown in Fig.~\ref{fig:4}, with bond dimension $\chi$ on all indices, which can be exactly contracted to a scalar $\mathcal Z_0$ with cost scaling as $O(\chi^6)$.

The network can be approximated as the sum of four networks with cost $O(\chi^5)$ using the PNE shown in Fig.~\ref{fig:4}(a). Note that the choice of partitioned indices is based on maximizing the degree of the residue terms, as discussed in the previous section. We also construct an $O(\chi^4)$ approximation using a different PNE based on six multi-index partitions (two vertical, two horizontal, and two diagonal), as shown in Fig.~\ref{fig:4}(b). Here we use the combinatorial form of the PNE in order to avoid the explicit use of complementary projectors $Q$, which would otherwise increase the cost scaling. It should be noted that Fig.~\ref{fig:4}(b) displays only the six principal network diagrams (i.e., those with a single active partition), while the sub-leading diagrams (those with multiple active partitions) are omitted for brevity. In this case, if $C(n,k)$ denotes the binomial coefficient, the total number of sub-leading networks is
\begin{equation}
C(6,2) + C(6,3) + C(6,4) + C(6,5) + C(6,6) = 57.
\end{equation}
Although evaluating this large collection of sub-leading networks may appear burdensome, it is typically manageable because (i) each term is computationally inexpensive to evaluate, with cost scaling only as $O(\chi^3)$, and (ii) their evaluation is straightforward to automate. Additionally, if the projectors $P$ are constructed from (products of) fixed-point BP messages, many of the expansion terms can be discarded automatically due to the BP fixed point forbidding any dangling excitations. Nevertheless, because the combinatorial PNE leads to an exponential proliferation of terms, it is generally unwise to employ a large number of distinct partitions within a single network. Fortunately, this exponential growth can be avoided through the use of a recursive partitioning strategy, as discussed in Sect.~\ref{sect:recursive}.

\begin{figure}[!t] 
\begin{center}
\includegraphics[width=8.5cm]{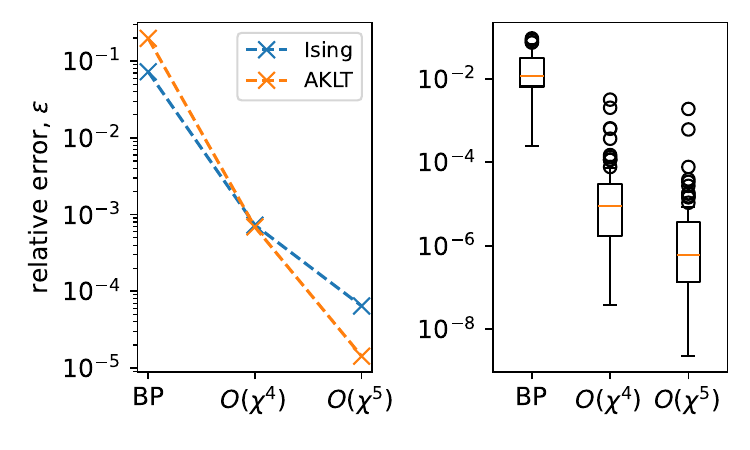}
\caption{Relative errors $\varepsilon$, as defined in Eq.~\ref{eq:B1}, for the contraction of the $3\times 3$ network shown in Fig.~\ref{fig:4}, comparing the BP approximation with the $O(\chi^4)$ and $O(\chi^5)$ partitioned network expansions. (left) Results obtained using tensors derived from either the $2D$ classical Ising model (at criticality) or the $2D$ square-lattice AKLT model, both blocked into networks with bond dimension $\chi=64$. (right) Aggregate results over 100 independent instances of random tensors with bond dimension $\chi=64$.}
\label{fig:4b}
\end{center}
\end{figure}

In Fig.~\ref{fig:4b} we benchmark the accuracy of these approximations to the $3\times 3$ network for (i) the $2D$ classical Ising model at criticality, (ii) the square-lattice AKLT model, and (iii) tensors populated with random elements. In all cases, we use tensor networks with bond dimension $\chi=64$, obtained by blocking lower-dimensional networks when necessary. For each example, we first compute a BP fixed point and then construct the projectors $P$ required for the various expansions from the fixed-point messages. These benchmarks show that the $O(\chi^4)$ PNE yields a substantial improvement over the base BP approximation, typically several orders of magnitude reduction in relative error, while the $O(\chi^5)$ expansion improves the accuracy further by at least an additional order of magnitude.

\begin{figure} [!t] 
\begin{center}
\includegraphics[width=8.5cm]{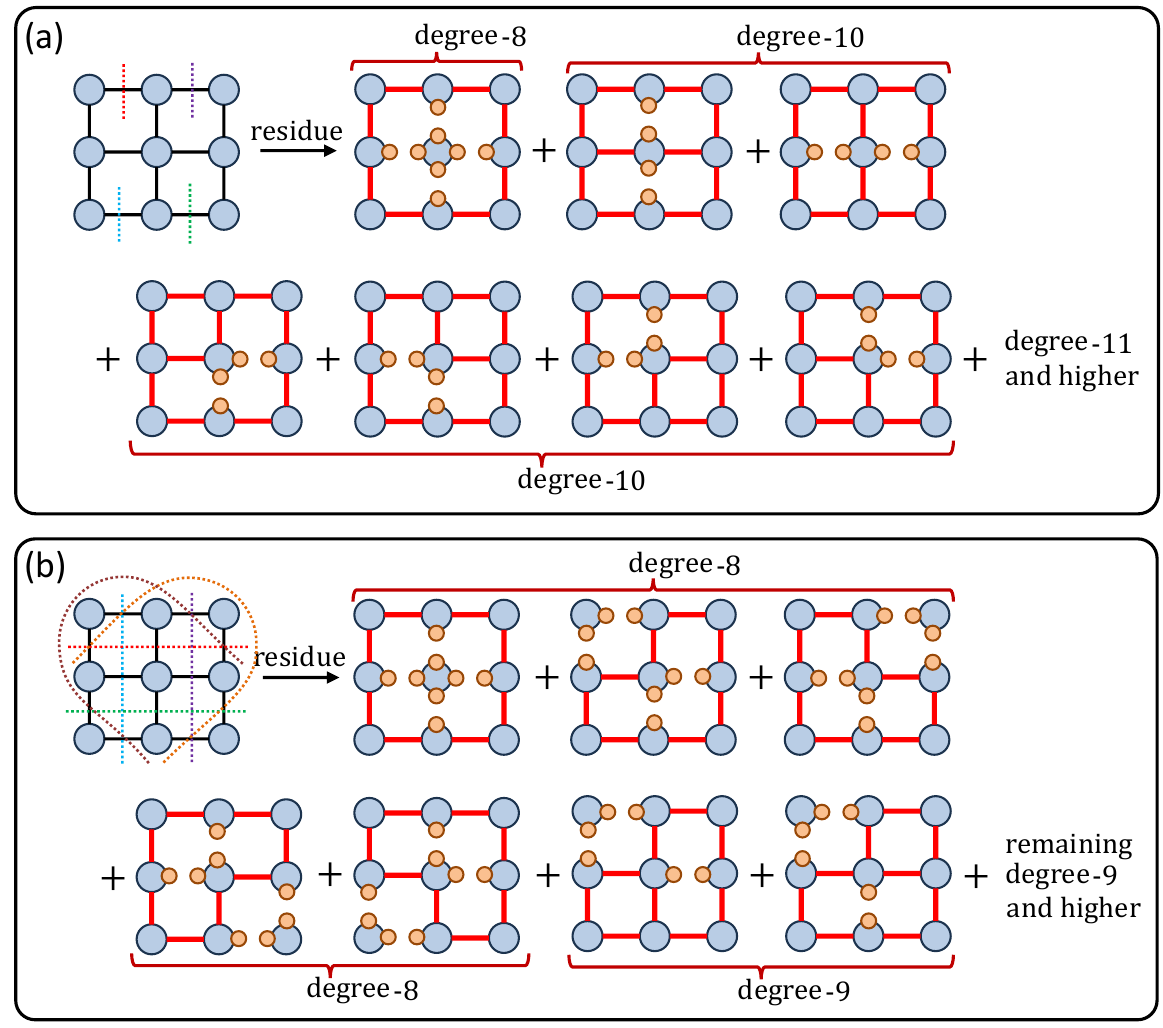}
\caption{(a) The low-degree residue terms from the expansion in Fig.~\ref{fig:4}(a). (b) Some of the low-degree residue terms from the expansion in Fig.~\ref{fig:4}(b).}
\label{fig:5}
\end{center}
\end{figure}

Once again, the relative accuracy of the different expansions can be understood by examining the components of the residue. The low-degree contributions to the residue from the $O(\chi^5)$ expansion are shown in Fig.~\ref{fig:5}(a); because the projectors $P$ were built from fixed-point BP messages, any residue terms containing dangling excitations have been discarded. The residue consists of a degree-8 term, six degree-10 terms, plus higher-degree terms. In contrast, the residue from the $O(\chi^4)$ expansion, shown in Fig.~\ref{fig:5}(b), contains five degree-8 terms as well as a large number of degree-9 and higher-degree terms. It therefore follows that the $O(\chi^4)$ expansion generally exhibits significantly larger errors than the $O(\chi^5)$ expansion.

\begin{figure}[!th] 
\begin{center}
\includegraphics[width=8.5cm]{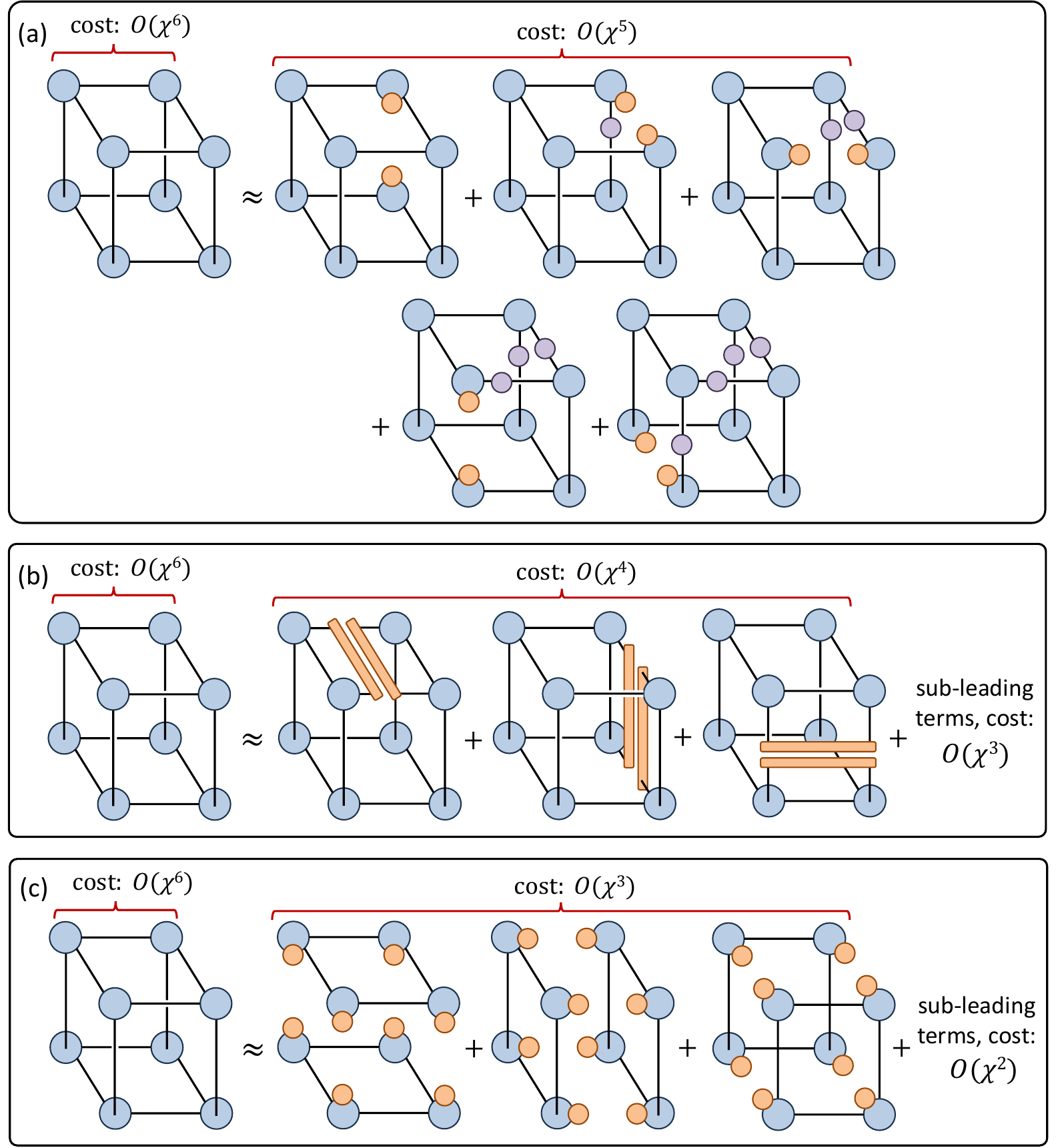}
\caption{(a) A network with exact contraction cost $O(\chi^6)$ is approximated as a sum of five cost $O(\chi^5)$ networks using a set of five single-index partitions. (b) The network is approximated as a sum of three cost $O(\chi^4)$ networks (plus sub-leading terms) using three double-index partitions. (c) The network is approximated as a sum of three cost $O(\chi^3)$ networks (plus sub-leading terms) using partitions that separate the network into products of $2D$ planes.}
\label{fig:6}
\end{center}
\end{figure}

\subsection{$2\times 2 \times 2$ Network}
We now consider an example involving the approximate contraction of a small $3D$ tensor network, as shown in Fig.~\ref{fig:6}. Assuming a bond dimension $\chi$ on all indices, this network can be exactly contracted to a scalar $\mathcal Z_0$ with cost scaling as $O(\chi^6)$. We construct an $O(\chi^5)$ approximation using a PNE with five single-index partitions, as illustrated in Fig.~\ref{fig:6}(a). Similarly, we obtain an $O(\chi^4)$ approximation by using the combinatorial form of the PNE together with a set of three double-index partitions, as shown in Fig.~\ref{fig:6}(b). Here we use rank $r=1$ projectors acting jointly on pairs of indices, without requiring these projectors to factorize into tensor products over individual indices. Finally, we construct an $O(\chi^3)$ approximation using the combinatorial PNE with three multi-index partitions, each of which acts to separate the network into a product of $2D$ planes; see also Fig.~\ref{fig:6}(c).

\begin{figure}[!t] 
\begin{center}
\includegraphics[width=8.5cm]{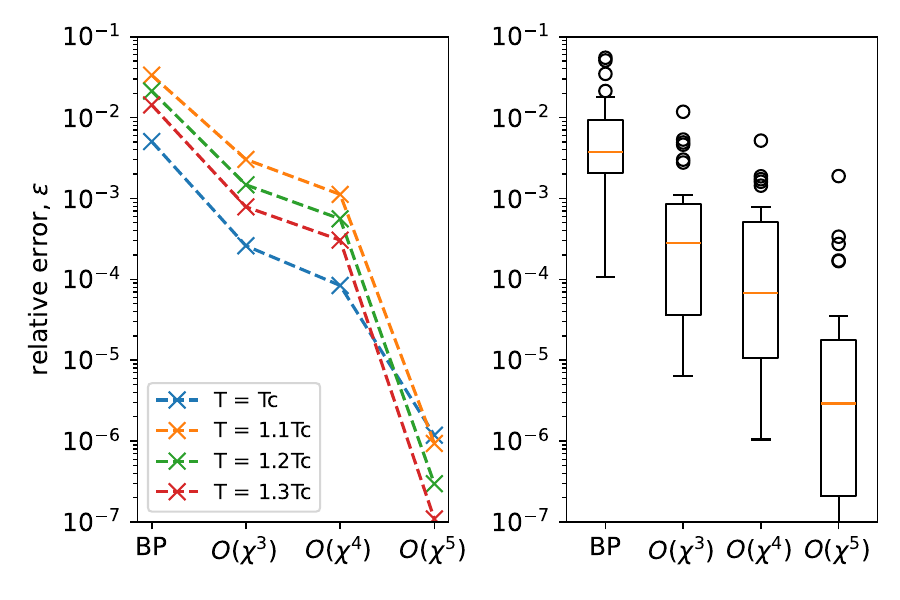}
\caption{Relative errors $\varepsilon$, as defined in Eq.~\ref{eq:B1}, for the contraction of the network in Fig.~\ref{fig:6}, comparing the BP approximation with the $O(\chi^3)$, $O(\chi^4)$, and $O(\chi^5)$ partitioned expansions. (left) Results obtained using tensors derived from the $3D$ classical Ising model, blocked into a bond-dimension $\chi=16$ network, at or near the critical temperature $T_C$. (right) Aggregate results over 30 independent instances of random tensors with bond dimension $\chi=16$.}
\label{fig:6b}
\end{center}
\end{figure}

We benchmark the accuracy of these approximations for the $2\times 2\times 2$ network using the $3D$ classical Ising model at various temperatures, as well as networks composed of random tensors; the results are shown in Fig.~\ref{fig:6b}. In all cases, the expansion is applied to networks with bond dimension $\chi=16$, which in the case of the Ising model is obtained via an initial coarse-graining step. As usual, we first compute a BP fixed point for each network and then construct the projectors $P$ from the fixed-point messages. However, for the $O(\chi^4)$ expansion in Fig.~\ref{fig:6}(b), we construct and employ two-site fixed-point BP messages obtained by grouping pairs of indices and running a separate message passing algorithm on this grouped structure. Across all test cases, we observe a systematic improvement in accuracy when progressing from the $O(\chi^3)$ to the $O(\chi^4)$ and then to the $O(\chi^5)$ expansions. This highlights a key strength of the partitioned expansion approach: it enables the construction of an approximation scheme tailored to a specific computational budget. Interestingly, the results in Fig.~\ref{fig:6b} show that the $3D$ Ising model is most challenging to approximate slightly above the critical temperature, at $T = 1.1\, T_C$. This contrasts with SVD-based approximations, which typically encounter their greatest difficulty at or extremely near criticality.

\subsection{$2\times 3$ Open Network}
\begin{figure} [!th] 
\begin{center}
\includegraphics[width=7.0cm]{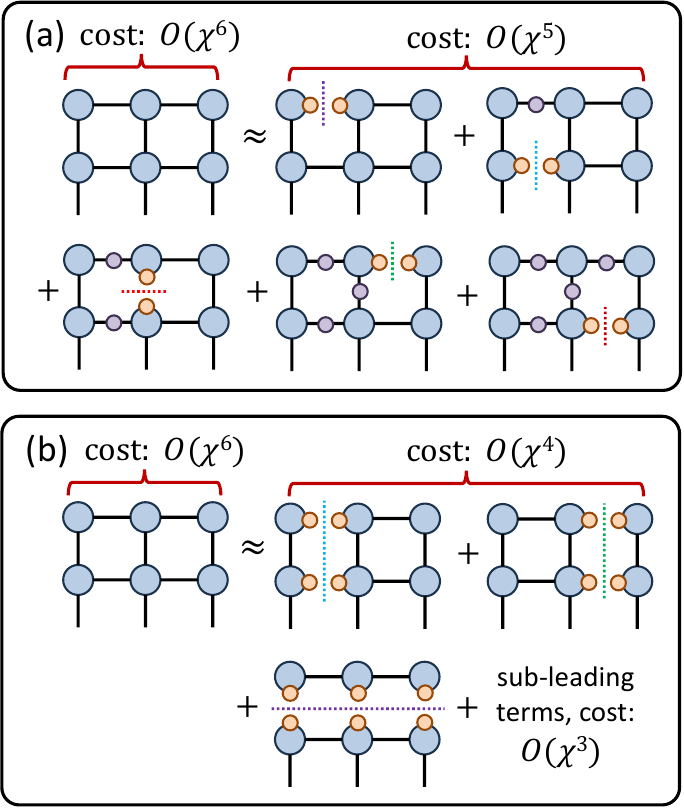}
\caption{(a) An open tensor network of contraction cost $O(\chi^6)$ is approximated as a sum of cost $O(\chi^5)$ networks. (b) The open tensor network is approximated as a sum of cost $O(\chi^4)$ networks (plus sub-leading terms).}
\label{fig:11}
\end{center}
\end{figure}

As a final example in this section, we consider the application of the PNE to evaluate the $2\times 3$ network shown in Fig.~\ref{fig:11}, which possesses several open indices. While evaluating closed networks is useful for extracting observables from a tensor network, many tensor network algorithms require the evaluation of open networks such as tensor environments, which are often needed for optimization algorithms. Thus, the ability of the PNE to evaluate open networks endows it with a broader range of potential applications than previous proposals restricted to scalar networks. Assuming that all indices have dimension $\chi$, the network in Fig.~\ref{fig:11} can be exactly contracted to a three-index tensor with cost scaling as $O(\chi^6)$. Using a set of single-index partitions, one can obtain an $O(\chi^5)$ expansion, as shown in Fig.~\ref{fig:11}(a). A cheaper $O(\chi^4)$ expansion can also be obtained through a set of multi-index partitions, as shown in Fig.~\ref{fig:11}(b).

\begin{figure}[!t] 
\begin{center}
\includegraphics[width=8.5cm]{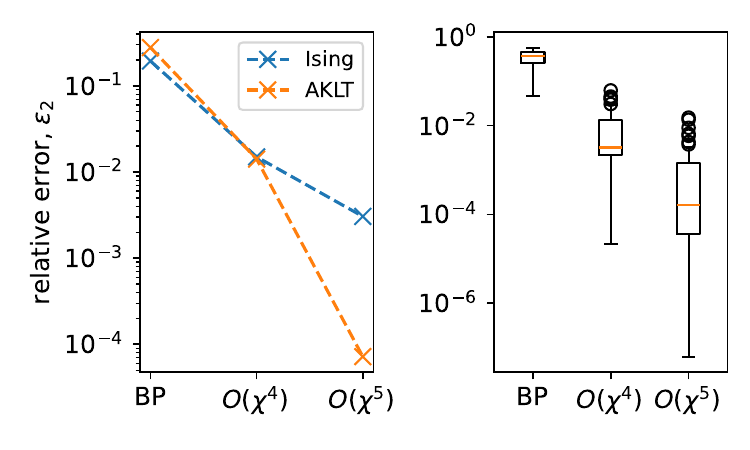}
\caption{Relative errors $\epsilon_2$, as defined in Eq.~\ref{eq:B2}, for the contraction of the network shown in Fig.~\ref{fig:11}. The base BP approximation is compared against the $O(\chi^5)$ and $O(\chi^4)$ expansions from Fig.~\ref{fig:11}(a) and Fig.~\ref{fig:11}(b), respectively. (Left) Results for the $2D$ classical Ising model (at criticality) and the $2D$ square-lattice AKLT model, each blocked into a bond-dimension $\chi=64$ network. (Right) Aggregate results from 100 samples of randomly generated tensors with bond dimension $\chi=64$.}
\label{fig:11b}
\end{center}
\end{figure}

We benchmark the accuracy of the open-network contractions for (i) the $2D$ classical Ising model at criticality, (ii) the square-lattice AKLT model, and (iii) randomly generated tensors. In all test cases, we work with networks of bond dimension $\chi=64$, obtained by performing a preliminary coarse-graining when necessary. As before, we first compute a BP fixed point for each network and then construct the projectors $P$ from the fixed-point messages. For the open network, we use a boundary-reflection approach to implement BP: the incoming message on each open index is set equal to the outgoing message on that index. If $T_\textrm{ex.}$ denotes the tensor resulting from exact contraction of the network, we compare the 2-norm error $\epsilon_2$ of the approximate tensors,
\begin{equation}
\epsilon_2 = \frac{\norm{T_\textrm{ex.} - T_\textrm{approx}}}{\norm{T_\textrm{ex.}}}.\label{eq:B2}
\end{equation}
The results, shown in Fig.~\ref{fig:11b}, are broadly consistent with the benchmarks for closed tensor networks: in every test case, the $O(\chi^4)$ expansion yields a substantial improvement in accuracy over the base BP approximation, while the $O(\chi^5)$ expansion improves the accuracy further.

\section{Larger Network Examples} \label{sect:large}

\subsection{Recursive Partitioning} \label{sect:recursive}
Although the partitioned network expansion was shown to be effective for the relatively small networks examined in Sect.~\ref{sect:small}, it is not immediately clear how to apply it in the same manner to larger networks. In particular, the combinatorial form of the PNE, which is required when employing multi-index partitions, contains $2^M$ terms, where $M$ is the number of partitions used. Thus, $M$ must remain relatively small, even though most of these terms are computationally inexpensive to evaluate. In general, it may be highly nontrivial to identify a small number $M$ of partitions that both (i) produce a small residue error in the expansion and (ii) lead to computationally cheap expansion terms.

In this section, we propose a workaround based on \emph{recursive} partitioning, which avoids the need to satisfy points (i) and (ii) simultaneously. The idea is straightforward: one begins by choosing a small number of partitions designed solely to minimize the residue (i.e., such that the residue contains only high-degree excitations) and then expands, but does not yet evaluate, the corresponding sub-networks. Any expansion term whose contraction cost exceeds the desired computational threshold is then individually re-partitioned and expanded into new sub-networks. This process is iterated, introducing progressively finer partitions, until all resulting sub-networks fall within the desired computational budget.

\begin{figure}[!t] 
\begin{center}
\includegraphics[width=8.0cm]{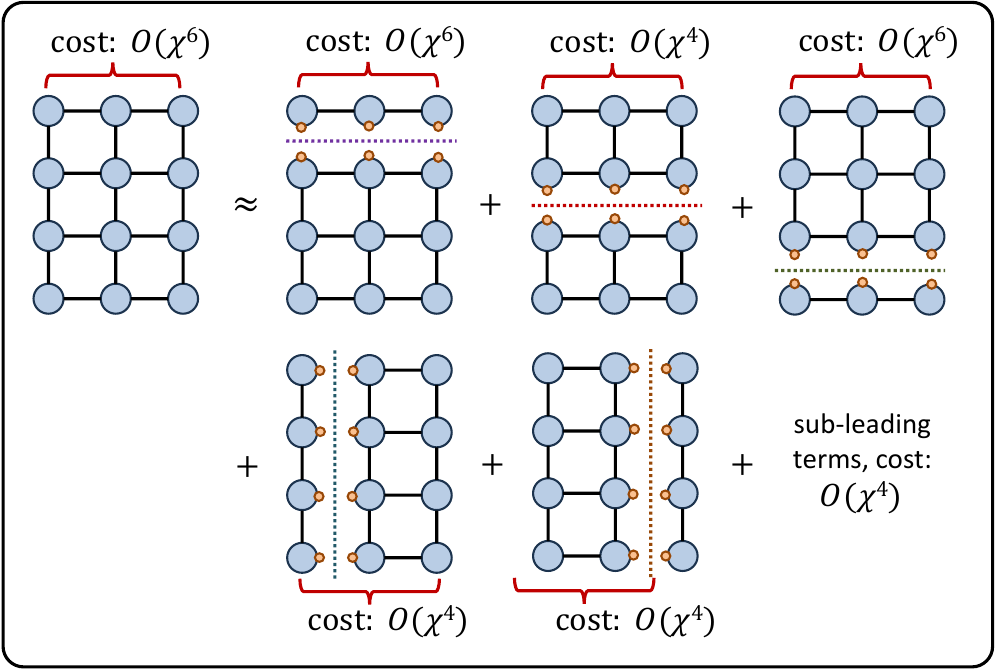}
\caption{The $4 \times 3$ square-lattice network, with an exact contraction cost $O(\chi^6)$, is partitioned into a sum of networks using a total of three horizontal and two vertical partitions. The resulting principal networks have contraction costs scaling as either $O(\chi^6)$ or $O(\chi^4)$. The $O(\chi^6)$ principal networks can be further approximated at $O(\chi^4)$ cost by applying the expansion from Fig.~\ref{fig:4}(b), yielding an overall $O(\chi^4)$ approximation to the original network.}
\label{fig:7}
\end{center}
\end{figure}

Consider the example $4\times 3$ network shown in Fig.~\ref{fig:7}, which has an exact contraction cost of $O(\chi^6)$. The most natural approach to expand would be to apply a set of vertical and horizontal partitions. However, as illustrated in Fig.~\ref{fig:7}, the resulting expansion still contains sub-networks corresponding to $3\times 3$ clusters, each of which has a contraction cost of $O(\chi^6)$. Hence, this expansion alone would not reduce the overall cost scaling relative to the original network. This issue can be resolved by applying an additional expansion to each $3\times 3$ cluster. In particular, the $O(\chi^4)$ expansion considered in Fig.~\ref{fig:4}(b) can be used to expand these clusters, thereby reducing the overall cost scaling of the full expansion to $O(\chi^4)$. 

Properly employed, this method of recursive partitioning could provide a systematic approach for approximating the contraction of completely arbitrary tensor networks, although a comprehensive exploration of this idea is left for future work.

\subsection{Comparison Against SVD Truncations}
In this section we compare the accuracy of approximate tensor network contraction using SVD truncation versus the partitioned expansion for a representative example network. In general, the accuracy of either approximation may depend on several factors, including: (i) the network geometry, (ii) the structure of the tensors within the network, and, most importantly, (iii) the specifics of how the approximation is implemented. Ref.~\onlinecite{evenbly2025loopseriesexpansionstensor} already presented an example in which a loop expansion was computationally cheaper than an SVD-based method (while producing equivalent accuracy), but only for networks composed specifically of so-called corner double-line (CDL) tensors. Our aim here is to present an example in which the PNE outperforms the SVD for network contraction without requiring any special choice of tensors.

\begin{figure}[!t] 
\begin{center}
\includegraphics[width=8.0cm]{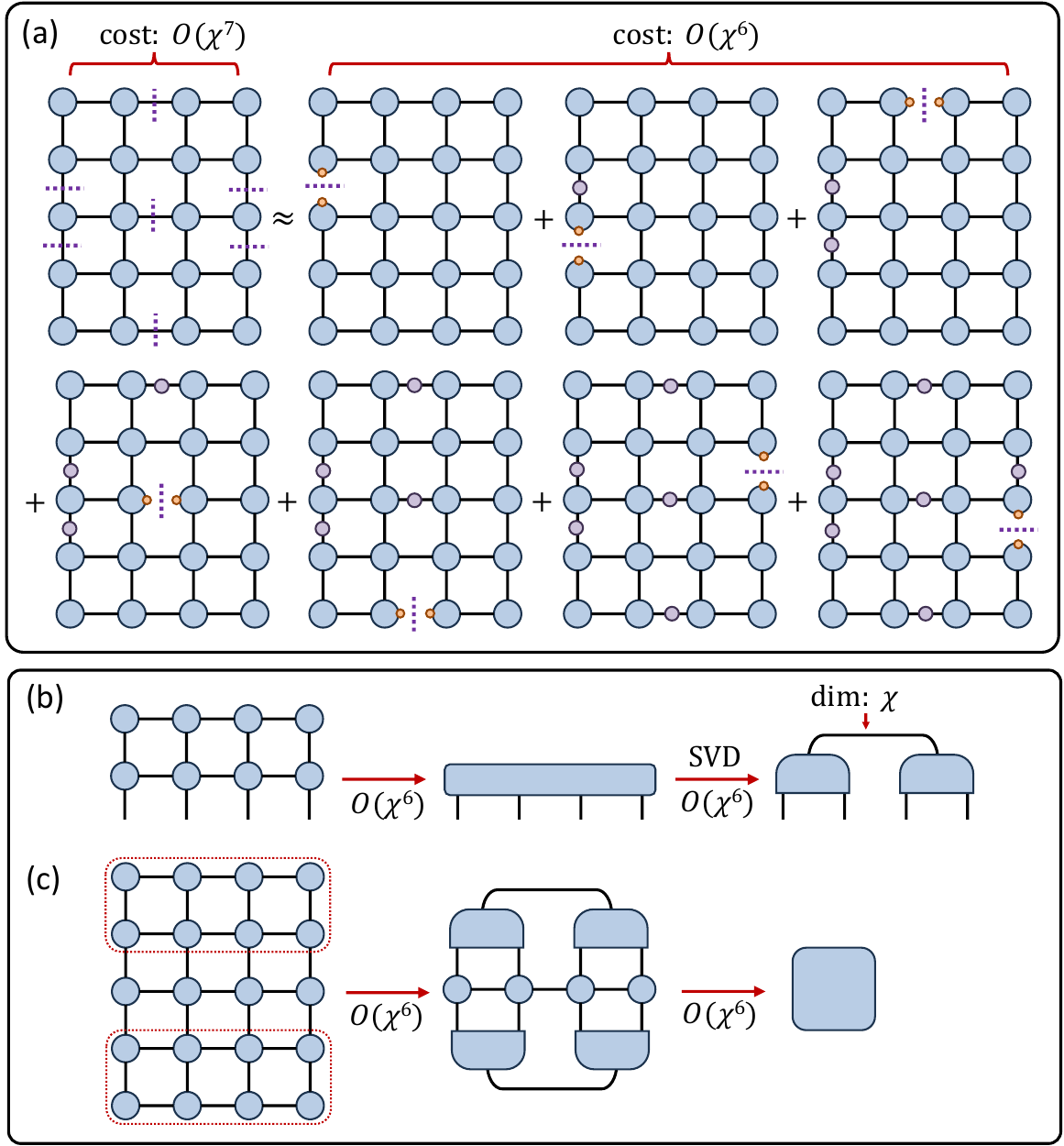}
\caption{(a) The $5 \times 4$ square-lattice network, whose exact contraction cost scales as $O(\chi^7)$, is approximated via a PNE as a sum of seven $O(\chi^6)$ networks. (b) A $2 \times 4$ region of the network can be contracted and then truncated into a rank-$\chi$ product of three-index tensors using the SVD. (c) The $5 \times 4$ network can be approximately contracted at cost $O(\chi^6)$ by using the SVD truncation shown in (b).}
\label{fig:9}
\end{center}
\end{figure}

In most applications, SVD-based truncations are used most effectively when decomposing a tensor with four (or more) indices into a product of tensors with three (or more) indices\cite{Kolda2009,gray2021hyper}. In this context, the SVD can be viewed as compressing multiple indices into a single effective index. Since the cost of decomposing a four-index tensor into a product of three-index tensors scales as $O(\chi^6)$ (assuming $\chi$-dimensional indices), such a truncation only reduces the overall contraction cost if the original network has contraction cost strictly greater than $O(\chi^6)$. Thus, it is not immediately obvious how the SVD could reduce the cost of any of the networks considered in Sect.~\ref{sect:small}, which all had exact contraction costs no larger than $O(\chi^6)$. Following this observation, we select as a test case the $5\times 4$ square-lattice tensor network shown in Fig.~\ref{fig:9}, which has an exact contraction cost of $O(\chi^7)$.

The network can be approximated using a partitioned expansion based on seven single-index partitions, as shown in Fig.~\ref{fig:9}(a), reducing the cost scaling to $O(\chi^6)$. The cost scaling can likewise be reduced to $O(\chi^6)$ using SVD truncations, as shown in Fig.~\ref{fig:9}(b--c). Here we follow a strategy similar to that employed in the corner transfer matrix method for contracting PEPS\cite{Orus2009Corner}: a four-index tensor is formed by exactly contracting the upper portion of the network, and this tensor is then decomposed into a product of three-index tensors while truncating to retain only the $\chi$ largest singular values. The network containing these truncated tensors can then be contracted with cost $O(\chi^6)$.

\begin{figure}[!t] 
\begin{center}
\includegraphics[width=8.5cm]{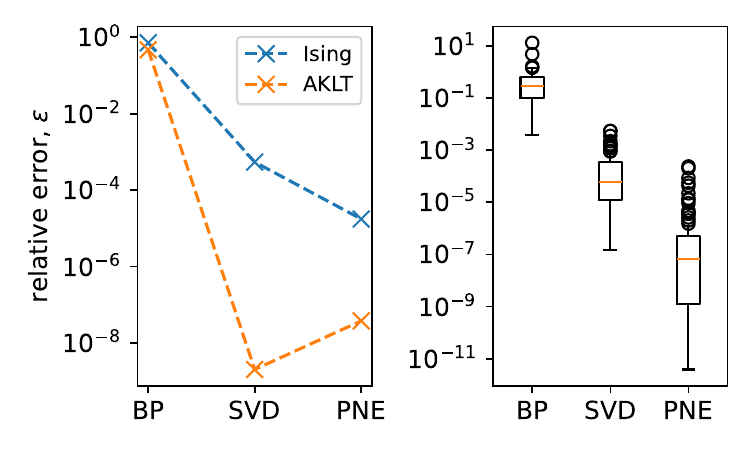}
\caption{Relative errors $\varepsilon$, as defined in Eq.~\ref{eq:B1}, for the contraction of the $5 \times 4$ network shown in Fig.~\ref{fig:9}. The BP approximation is compared with both the $O(\chi^6)$ SVD-based approximation from Fig.~\ref{fig:9}(c) and the $O(\chi^6)$ PNE from Fig.~\ref{fig:9}(a). (Left) Results for the $2D$ classical Ising model (at criticality) and the $2D$ square-lattice AKLT model, each blocked into bond-dimension $\chi=16$ networks. (Right) Aggregate results from 100 samples of randomly generated tensors with bond dimension $\chi=16$.}
\label{fig:9b}
\end{center}
\end{figure}

The accuracy of the different approximation strategies for the $5\times 4$ network is compared numerically in Fig.~\ref{fig:9b}. This comparison includes networks derived from (i) the $2D$ classical Ising model, (ii) the square-lattice AKLT model, and (iii) random tensors. In all cases, the networks have bond dimension $\chi=16$, obtained via preliminary blocking in cases (i) and (ii). The results show that the PNE yields a significantly smaller approximation error than the SVD-based truncations, often by several orders of magnitude, for both the Ising and random networks. However, for the specific case of the AKLT network, the SVD truncation is seen to produce the more accurate result.

Finally, we remark that although this comparison provides useful insight into the relative performance of the methods, it is not necessarily appropriate to view the two approximation strategies as being in direct competition. In practice, SVD-based truncations and partitioned expansions may be used together, and in a complementary fashion, within more sophisticated tensor-network algorithms.

\subsection{Infinite Tensor Networks}
While the recursive partitioning strategy discussed in Sect.~\ref{sect:recursive} allows one to treat larger finite networks, it remains unclear how the PNE could be used to evaluate \emph{infinite} networks. In this section we describe a strategy for applying the PNE to an infinite, translation-invariant, square-lattice tensor network composed of uniform tensors. This approach can be generalized straightforwardly to the case of a single impurity tensor—relevant, for example, when computing expectation values in an infinite PEPS—although we do not consider that setting explicitly here.

\begin{figure}[!t] 
\begin{center}
\includegraphics[width=8.5cm]{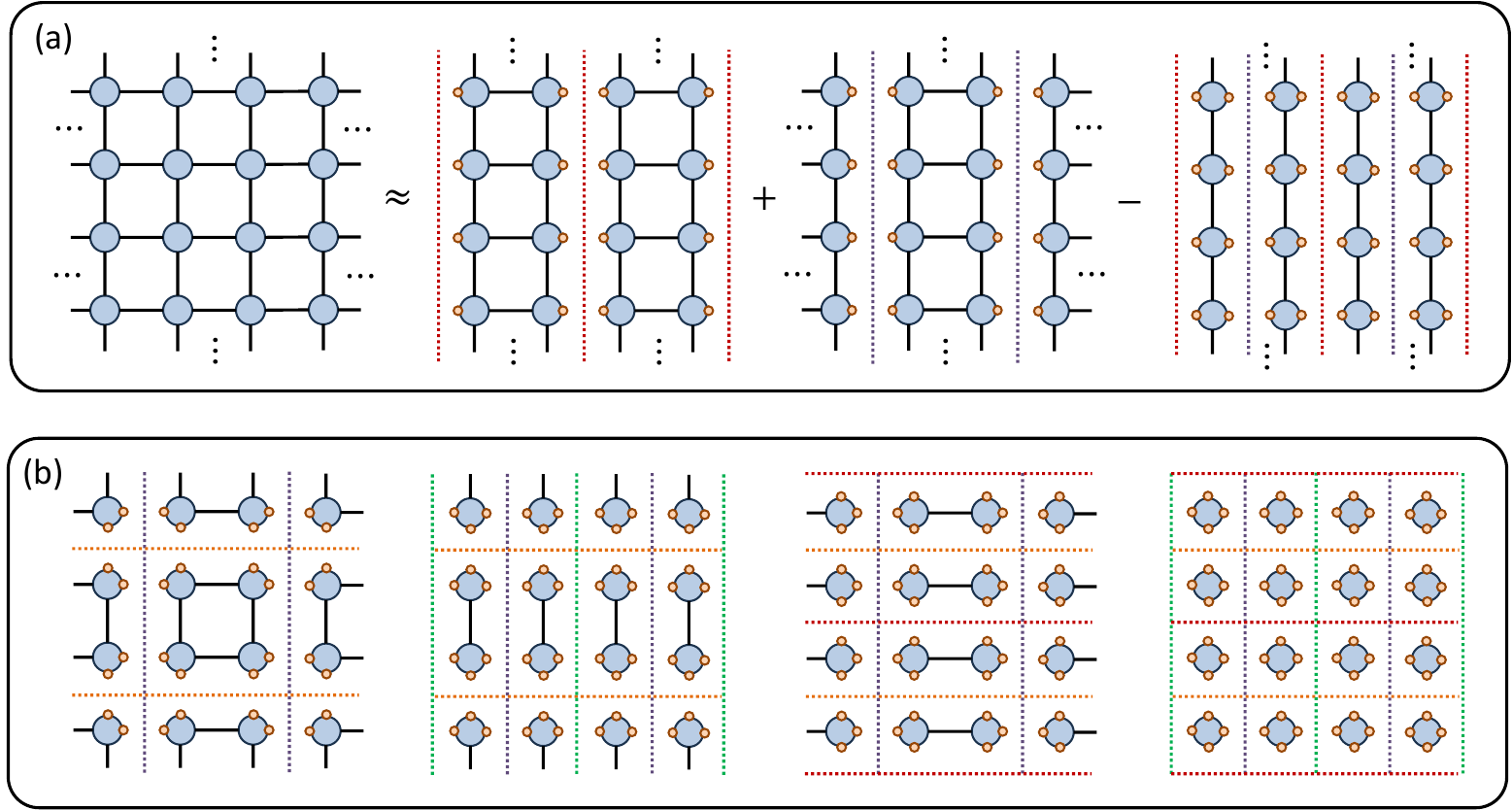}
\caption{(a) A partitioned expansion for the infinite square-lattice network, obtained by partitioning the lattice into vertical width-2 strips using either an odd–even or even–odd pattern. (b) A depiction of the sub-leading terms generated when additional partitions formed from horizontal width-2 strips are included in the PNE from (a).}
\label{fig:12}
\end{center}
\end{figure}

A natural but simple approach would be to isolate a finite patch of $N$ sites by fixing its boundary indices using fixed-point BP messages, and then apply the PNE to the resulting finite network. Although viable, we pursue a more sophisticated method. Suppose that our goal is to evaluate the free energy density,
\begin{equation}
f = -\lim_{N\rightarrow\infty} \frac{\log \left( \mathcal Z_N \right)}{N}.
\end{equation}
A first approximation can be obtained by partitioning the infinite network into semi-infinite width-2 strips, yielding
\begin{equation}
f \approx -\frac{\log(\lambda_2)}{2},
\end{equation}
where $\lambda_2$ is the leading eigenvalue of the width-2 transfer matrix. In this setting, we envision the set of vertical slices as comprising a single partition, such that the expansion consists of only a single term. A more accurate approximation results from partitioning the lattice into two alternating sets of width-2 strips, following either an odd–even or even–odd pattern, as illustrated in Fig.~\ref{fig:12}(a). In this case,
\begin{equation}
f \approx -\frac{\log\!\left(2\lambda_2 - (\lambda_1)^2\right)}{2},
\end{equation}
where $\lambda_1$ is the leading eigenvalue of the width-1 transfer matrix, which arises from consideration of the sub-leading term in the expansion.
One may similarly partition the lattice into three sets of width-3 strips, giving
\begin{equation}
f \approx -\frac{\log\!\left(3\lambda_3 + 3\lambda_2 \lambda_1 + (\lambda_1)^3\right)}{3},
\end{equation}
with $\lambda_3$ the leading eigenvalue of the width-3 transfer matrix. In principle this process can be extended to build an $L$-strip expansion of $f$, provided that the eigenvalues $\{\lambda_1,\ldots,\lambda_L\}$ can be computed. Rather than partitioning along only one direction, one can simultaneously partition into vertical and horizontal strips. This reduces the residue by capturing longer-range excitations along both axes. The expansion then contains sub-leading terms involving both horizontal and vertical partitions, as depicted in Fig.~\ref{fig:12}(b), which reduce to products of finite disconnected patches. Using width-2 vertical and horizontal strips yields
\begin{equation}
f \approx -\frac{\log\!\left(R_1 - R_2 + R_3 - R_4\right)}{4},
\end{equation}
where the $R_i$ denote expansion terms of order $i$:
\begin{align}
R_1 &= 2(\lambda_2)^2 + 2(\gamma_2)^2, \nonumber \\
R_2 &= 4\mathcal{Z}_{22} + (\lambda_1)^4 + (\gamma_1)^4, \nonumber \\
R_3 &= 2(\mathcal Z_{21})^2 + 2(\mathcal Z_{12})^2, \nonumber \\
R_4 &= (\mathcal Z_{11})^4.
\end{align}
Here $\lambda_k$ and $\gamma_k$ are the leading eigenvalues of the width-$k$ vertical and horizontal transfer matrices, while $\mathcal Z_{kp}$ denotes the partition function of a finite $k \times p$ patch. This approach generalizes naturally to width-$L$ strips, assuming the transfer-matrix eigenvalues $\{\lambda_k\}$ and $\{\gamma_k\}$ (for $k\le L$) and the patch contractions $\mathcal Z_{kp}$ (for $k,p\le L$) can be computed.

\begin{figure}[!th] 
\begin{center}
\includegraphics[width=8.5cm]{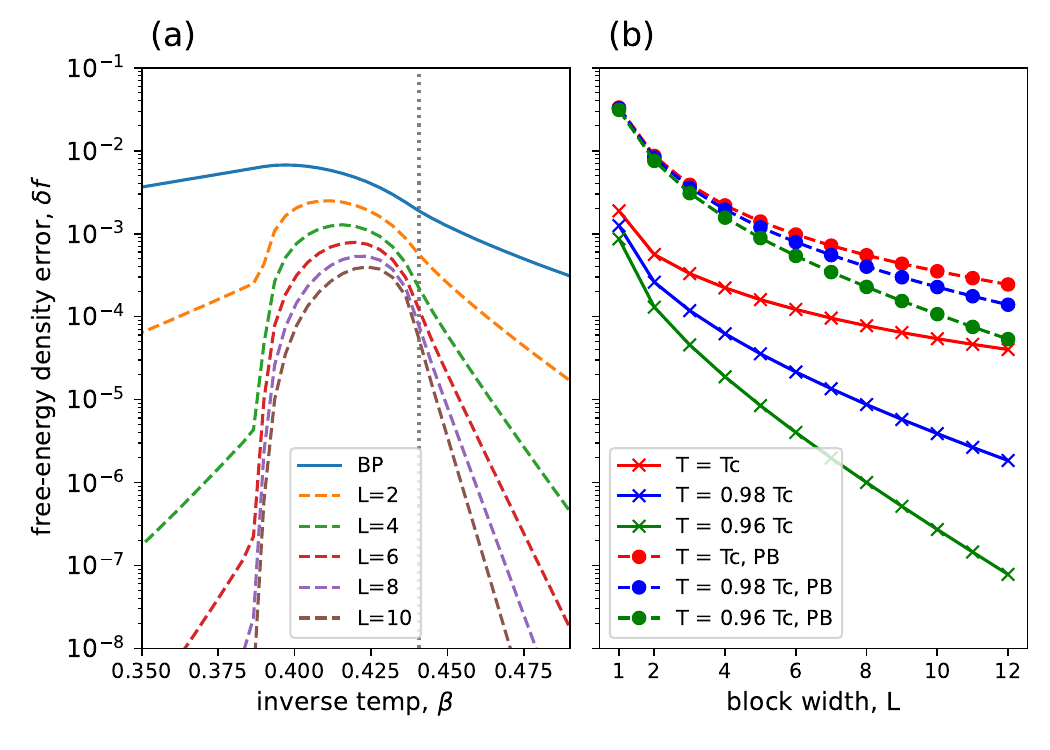}
\caption{Error in the free energy density for the infinite $2D$ classical Ising model, using expansions based on partitioning the network into width-$L$ horizontal and vertical strips. (a) Error as a function of inverse temperature $\beta$, with the vertical dotted line indicating the critical temperature $\beta_C \approx 0.4407$. (b) Solid lines show error from the partitioned expansions using width-$L$ strips, while dashed lines show error obtained from exact contraction of the $2D$ Ising model on an infinite-length cylinder of circumference $L$.}
\label{fig:12b}
\end{center}
\end{figure}

We benchmark this expansion on the infinite $2D$ classical Ising model. After a preliminary $2\times 2$ blocking step, the tensor network has bond dimension $\chi=4$. We compute the free energy density $f$ using PNE expansions with strip width up to $L=12$, corresponding to $24$ Ising spins. As shown in Fig.~\ref{fig:12b}, all PNE expansions provide substantial improvement over the BP approximation, with larger $L$ yielding systematically higher accuracy. Figure~\ref{fig:12b}(a) shows that the expansions perform best far from the critical temperature $T_C$, and, somewhat unexpectedly, are least accurate slightly above criticality at $T \approx 1.05 T_C$.

In Fig.~\ref{fig:12b}(b), the PNE results are compared against exact diagonalization of the transfer matrix on a cylinder of circumference $L$, where the results from the partitioned expansions are seen to be significantly more accurate. In fact, a circumference-$3L$ cylinder is required to roughly match the accuracy of an $L$-width PNE, even though both methods have similar computational cost for the same $L$. 

These results highlight a key advantage over previous strategies\cite{evenbly2025loopseriesexpansionstensor,midha2025beliefpropagationclustercorrectedtensor} for improving BP by summing individual loop  corrections: summing all of the individual correction terms that fit within a width-$12$ strip, which would otherwise be necessary to match the accuracy of Fig.~\ref{fig:12b}, would be infeasible due to the proliferation of such terms.

\begin{figure}[!th] 
\begin{center}
\includegraphics[width=7.0cm]{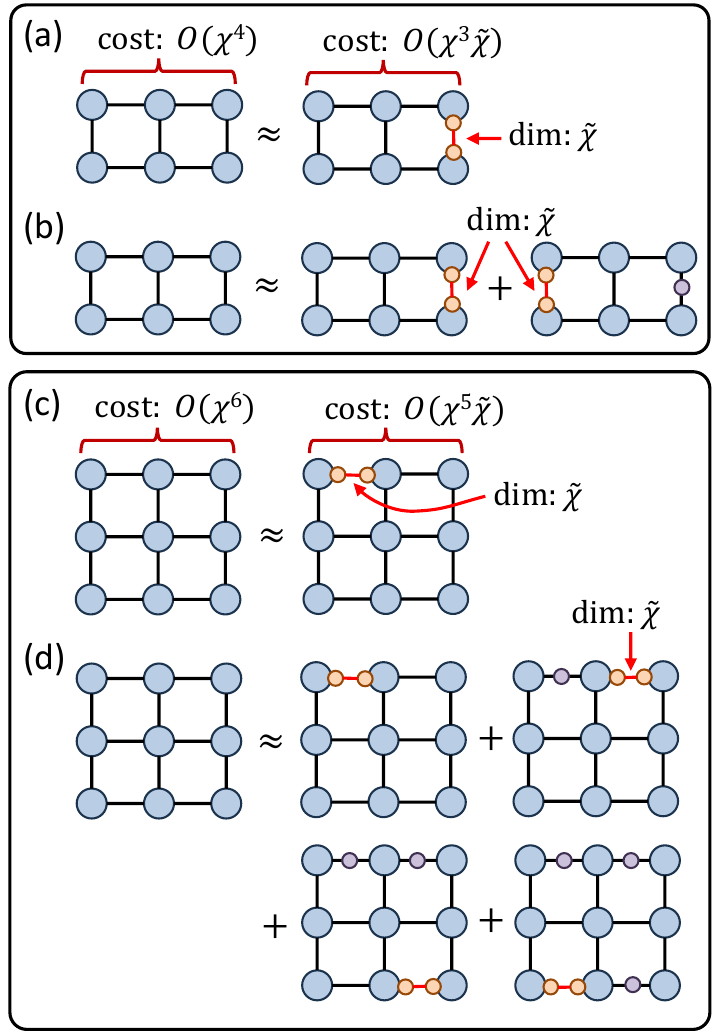}
\caption{(a) The $2 \times 3$ tensor network is approximated by truncating a single index from dimension $\chi \rightarrow \tilde{\chi}$ using a single rank-$\tilde{\chi}$ projector. (b) The same network approximated using a partitioned expansion with two rank-$\tilde{\chi}$ projectors. (c) The $3 \times 3$ network approximated by truncating a single index from dimension $\chi \rightarrow \tilde{\chi}$ using a single rank-$\tilde{\chi}$ projector. (d) The same network approximated using a partitioned expansion with four rank-$\tilde{\chi}$ projectors.}
\label{fig:10}
\end{center}
\end{figure}

\section{Beyond BP-based Approximations} \label{sect:beyond}
\subsection{Higher-rank Projections}
The examples considered thus far have focused on expansions based on rank $r=1$ projectors $P$ formed from fixed-point BP messages. In this section we explore expansions that employ higher-rank projectors. The previous examples demonstrated that different partitioning strategies can be used to adjust the contraction-cost scaling to fit a desired computational budget. However, tuning the projector rank provides an alternative means of controlling the accuracy of the expansion, one that allows direct refinement arbitrarily close to the exact result. Since BP messages intrinsically describe only a rank $r=1$ subspace, a different strategy is required to construct higher-rank projectors $P$. In Appendix~\ref{sect:weight} we describe a \emph{weight passing} algorithm in which, instead of assigning an incoming/outgoing message pair to each index $i$, a diagonal matrix $S_i$ of positive weights is obtained (which can be interpreted as a collection of multiple distinct incoming/outgoing message pairs weighted by their likelihood).

As a test of higher-rank projectors, we revisit the $2\times 3$ and $3\times 3$ networks studied earlier in Sect.~\ref{sect:small}. We construct networks from random tensors of bond dimension $\chi=64$, and the weight passing algorithm is iterated until the index weights $S_i$ are sufficiently converged for each index $i$. A rank-$\tilde{\chi}$ projector $P_i$ is then formed from the subspace associated with the largest weights in $S_i$. We consider two types of approximations: one based on a single rank-$\tilde{\chi}$ projector, and another based on a partitioned expansion using multiple rank-$\tilde{\chi}$ projectors, as illustrated in Fig.~\ref{fig:10}.

For the $2\times 3$ network, both approximations reduce the leading contraction cost from $O(\chi^4)$ to $O(\chi^3 \tilde{\chi})$, although the multi-partition approach is twice as expensive because two networks must be contracted. Similarly, for the $3\times 3$ network, both approximations reduce the leading cost from $O(\chi^6)$ to $O(\chi^5 \tilde{\chi})$, but the multi-partition expansion is four times as expensive due to the need to contract four networks in the expansion.

\begin{figure}[!th] 
\begin{center}
\includegraphics[width=8.5cm]{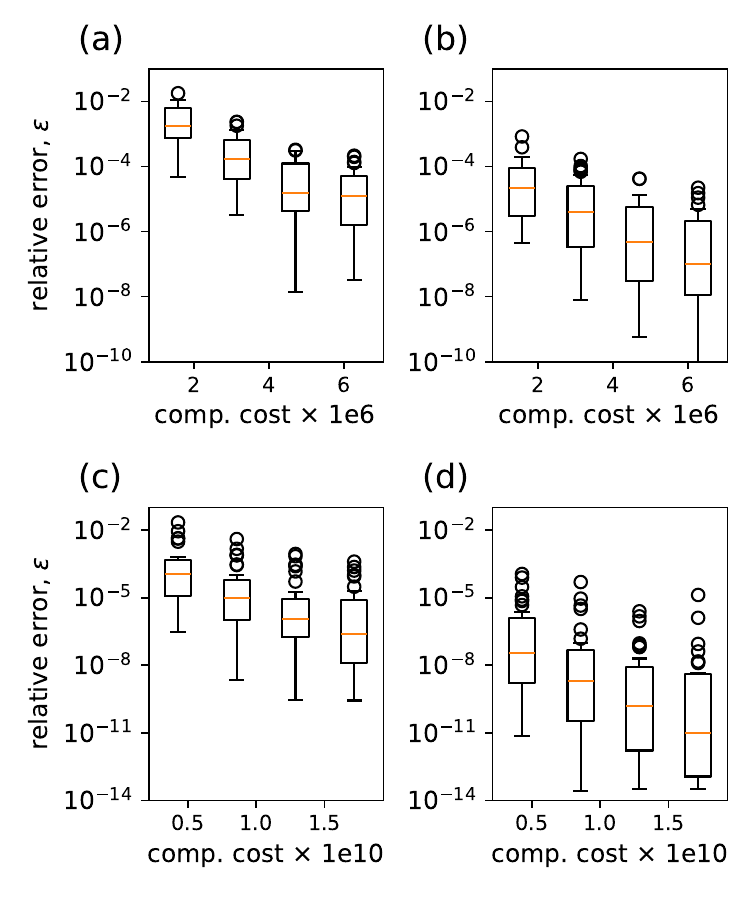}
\caption{Relative errors $\varepsilon$, as defined in Eq.~\ref{eq:B1}, plotted as a function of total computational cost for the expansions of Fig.~\ref{fig:10}, applied to 30 trials of networks composed of random tensors with bond dimension $\chi=64$. (a) Errors for the approximation of Fig.~\ref{fig:10}(a), using a single projector of rank $\tilde{\chi} = \{2,4,6,8\}$. (b) Errors for the partitioned expansion of Fig.~\ref{fig:10}(b), using projectors of rank $\tilde{\chi} = \{1,2,3,4\}$. (c) Errors for the approximation of Fig.~\ref{fig:10}(c), using a single projector of rank $\tilde{\chi} = \{4,8,12,16\}$. (d) Errors for the partitioned expansion of Fig.~\ref{fig:10}(d), using projectors of rank $\tilde{\chi} = \{1,2,3,4\}$.}
\label{fig:10b}
\end{center}
\end{figure}

In Fig.~\ref{fig:10b} we compare the accuracy of these approximations as a function of the total computational cost. For a fair comparison, smaller values of $\tilde{\chi}$ are chosen for the multi-index expansions so that their total costs match those of the single-projector approximations. We find that all approximations systematically converge toward the exact result as the projector rank $\tilde{\chi}$ increases. However, those based on multiple partitions converge more rapidly (as a function of total computational cost) than those based on a single partition. These examples also demonstrate that partitioned expansions can remain effective even without access to a BP fixed point.

\subsection{Degenerate Fixed Points} \label{sect:degen}
Previous proposals for introducing corrections to BP\cite{Chertkov2006LoopCalculus,Chertkov2006LoopSeries,evenbly2025loopseriesexpansionstensor,park2025simulatingquantumdynamicstwodimensional,midha2025beliefpropagationclustercorrectedtensor,gray2025tensornetworkloopcluster} would generally fail to produce accurate results when applied to networks with degenerate (or nearly degenerate) BP fixed points, since their validity relies on incorporating corrections only in the vicinity of a single BP fixed point. In this section we demonstrate how partitioned expansions can overcome this limitation, using the classical Ising model at sub-critical temperature as an illustrative example.

We begin by constructing a tensor-network representation of the $2D$ classical Ising model at temperature $T$ on a $12 \times 12$ lattice with open boundaries. This network is then coarse-grained into a $3 \times 3$ network of bond dimension $\chi=16$ by grouping $4 \times 4$ blocks of Ising spins. As a first test, we solve for a BP fixed point of the coarse-grained network and apply both the $O(\chi^4)$ and $O(\chi^5)$ partitioned expansions described in Sect.~\ref{sect:three}, using projectors $P$ of rank $\tilde{\chi}=1$ . We then apply the weight passing algorithm of Appendix~\ref{sect:weight} to obtain fixed-point weight matrices $S_i$ for each index $i$ in the network. Using these weights, we construct projectors $P$ of rank $\tilde{\chi}=2$  by selecting the subspace associated with the two largest weights from each $S_i$, and apply the same partitioned expansions using these higher-rank projectors.

\begin{figure}[!t] 
\begin{center}
\includegraphics[width=8.5cm]{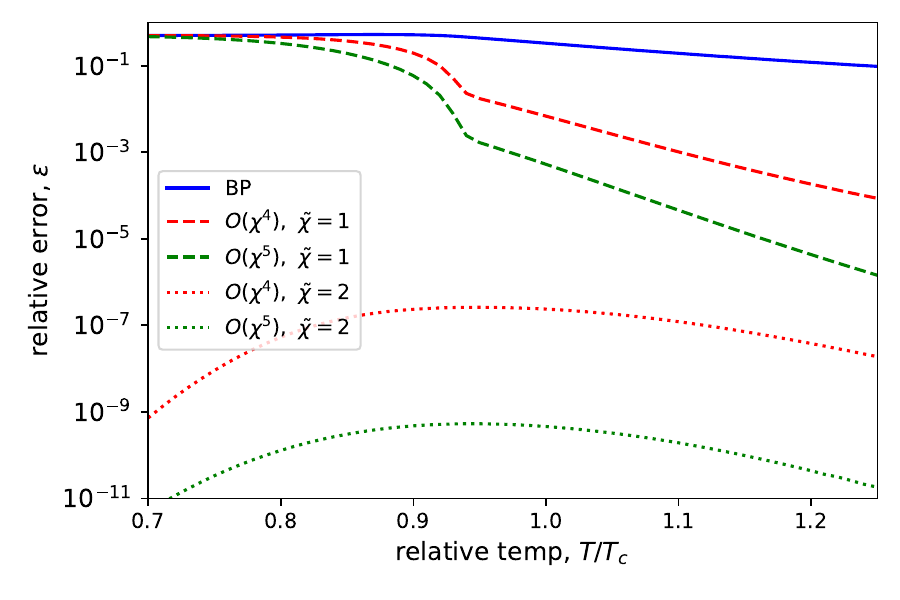}
\caption{Relative errors $\varepsilon$, as defined in Eq.~\ref{eq:B1}, for the contraction of the $2D$ classical Ising model on a $12 \times 12$ lattice with open boundaries, plotted as a function of temperature $T$. The model is first coarse-grained into a $3 \times 3$ network of bond-dimension $\chi=16$ tensors, then evaluated using either the BP approximation or the $O(\chi^5)$ and $O(\chi^4)$ PNE schemes of Fig.~\ref{fig:4}(a)--(b). In the low-temperature regime, an expansion using projectors $P$ of rank $\tilde{\chi}=2$ is required to obtain an accurate approximation.}
\label{fig:10c}
\end{center}
\end{figure}

The numerical results of this comparison are shown in Fig.~\ref{fig:10c}. We observe that the BP approximation, as well as the $\tilde{\chi}=1$ expansions built around the BP fixed point, fail dramatically in the low-temperature regime: they produce nearly $50\%$ relative error in the evaluation of $\mathcal Z_0$ for $T < 0.8\, T_C$, where $T_C$ is the critical temperature. This failure can be attributed to the fact that, at low temperatures, the model possesses two distinct BP fixed points corresponding to the $(+)$-spin-dominated and $(-)$-spin-dominated sectors. Expanding around only one of these fixed points is insufficient, as it necessarily omits the contribution from the other sector. In contrast, the expansions of rank $\tilde{\chi}=2$ incorporate contributions from both spin sectors and consequently yield extremely small relative errors, with $\varepsilon < 10^{-7}$ across the entire temperature range investigated.

Although this is a simple example, it illustrates clearly how BP-based methods can break down in the presence of degenerate fixed points, and how partitioned expansions circumvent this issue. As an alternative, we also explored performing separate rank-$\tilde{\chi}=1$ expansions around each of the $(+)$-spin and $(-)$-spin BP fixed points, then summing the resulting contributions. This approach also proved unsatisfactory, as the two BP fixed points are only strictly orthogonal in the zero-temperature limit. At any finite temperature, attempting to sum the results from the two BP fixed point expansions leads to systematic error due to an effective double counting of certain contributions.

\section{Discussion} \label{sect:discussion}
In this manuscript we introduced a general approach for approximating tensor-network contractions. Our proposed method can, in principle, be applied to the contraction of essentially any tensor network, including both finite and infinite networks, closed networks and networks with open indices, as well as networks for which BP-based methods fail due to degeneracy. The approach is also highly flexible: by adjusting how a network is partitioned, one can easily tune the tradeoff between accuracy and computational cost, with more accurate results obtained by using partitions that are shorter and more numerous. We expect that partitioned expansions may serve as a useful complement to SVD-based approximations, since the two methods appear to have different regimes of effectiveness. Indeed, for many of the example networks that were used to demonstrate the PNE, it is not obvious how SVD truncations could even be applied to reduce the contraction cost. Partitioned expansions may be especially advantageous for networks that are challenging for SVD-based methods, such as $3D$ networks\cite{Vlaar2021} and unstructured networks\cite{gray2024hyperoptimized}.

Our proposal also circumvents several failure modes of BP-based approaches. In particular, the method does not require a known BP fixed point to be applicable, and it can still produce accurate results in the presence of degenerate BP fixed points. However, for the expansion to be accurate, the dominant-weight subspace associated with each network index must still be identified. Appendix~\ref{sect:weight} introduced an iterative weight passing algorithm capable of producing these index weights, including for some networks in which BP fails to converge. Nonetheless, the weight passing method can also fail to identify a dominant subspace for difficult tensor networks. Examples include networks where tensors entries are drawn uniformly from the interval $[-1,1]$ without any bias, which are known to be more challenging than networks with a positive bias\cite{Chen2025Sign}. In such cases the resulting weights $S$ are nearly flat, and expansions based on low-rank projectors $P$ constructed from these weights yield poor accuracy. The convergence properties of weight passing, its relationship to BP, and more sophisticated variants will be explored in future work.

Although not a primary focus of this manuscript, we remark that partitioned expansions are relatively straightforward to implement numerically, especially when using a dedicated tensor-contraction library\cite{pfeifer2015ncontensornetworkcontractor,gray2018quimb,TensorOperations,itensor}. This is because all terms in the expansion generally share the same network geometry as the original network and therefore do not require separate coding. Consider an index $i$ of dimension $d$ connecting two tensors $A$ and $B$, and suppose a rank-$r$ projector $P$ is inserted on this index, represented as the outer product of a $(d \times r)$ isometry with its conjugate. By absorbing the isometries into the adjacent tensors, one obtains new tensors $\tilde{A}$ and $\tilde{B}$ connected by an index $i'$ of dimension $r$. Thus, each expansion term can be viewed as having the same network geometry but with modified tensors. This contrasts with previous approaches\cite{Chertkov2006LoopCalculus,Chertkov2006LoopSeries,evenbly2025loopseriesexpansionstensor,midha2025beliefpropagationclustercorrectedtensor}, where individual terms in the expansion typically correspond to separate networks with distinct geometries. The rules described in Eq.~\ref{eq:A4} and Eq.~\ref{eq:A6} for constructing the linear and combinatorial forms of the expansion are also straightforward to automate. Accordingly, it is possible to augment existing network-contraction routines to automatically compute the partitioned expansion when provided with the desired set of partitions and associated projectors $P$. Such an automated routine was used to perform the benchmark calculations in this work and will be described in more detail in future work.

\appendix

\section{Construction of Benchmark Examples} \label{sect:construction}
In this appendix we provide details on the construction of the various networks used in the benchmarking examples.

\subsection{Ising Tensors}
Many of the benchmark networks considered are derived from the $2D$ and $3D$ classical Ising model. In these cases, we begin by constructing a $\chi=2$ tensor network for the infinite square or cubic lattice using a standard encoding\cite{Levin2007TRG} of the partition function at temperature $T$. For the purposes of providing more challenging test cases, it is often helpful to perform a preliminary coarse-graining step to obtain tensor networks of larger bond dimension, such as $\chi=16$ or $\chi=64$, by grouping and reshaping local patches into single larger tensors.

Unless noted otherwise, we solve for the BP fixed-point messages of the infinite network and then construct the finite networks used in the benchmarks (e.g., the $3\times 3$ patch of Fig.~\ref{fig:4}) by inserting the fixed-point messages around the boundary of an appropriately sized finite patch. This method of constructing finite networks has two practical advantages:  
(i) it yields networks in which the fixed-point BP messages are homogeneous throughout, and  
(ii) it allows the same tensors and messages to be reused across many of the benchmark examples.

We also tested constructing finite networks directly from the partition function of the Ising model on a finite lattice with open boundary conditions. In practice, this produced results that were broadly comparable to those obtained from the infinite network using the boundary-insertion method described above.

\subsection{AKLT Tensors}
Some benchmark networks were derived from the square-lattice AKLT model\cite{Affleck1987AKLT,Wei2015AKLT2D}. In this setting, we began by constructing a square-lattice PEPS of local dimension $d=5$ and bond dimension $m=2$ for the exact AKLT ground state $\ket{\psi_{\mathrm{AKLT}}}$ using the standard construction. The scalar product $\braket{\psi_{\mathrm{AKLT}}}{\psi_{\mathrm{AKLT}}}$ is then formed by contracting the bra and ket PEPS over their physical indices, yielding a square-lattice tensor network (with no open indices) of bond dimension $\chi = m^2$.

As with the Ising model, it was useful to perform preliminary coarse-graining to obtain a tensor network of bond dimension $\chi = 16$ or $\chi = 64$ by grouping and reshaping tensors. The infinite network is then reduced to the finite benchmark networks using the same boundary-insertion method employed for the Ising tensors.

\subsection{Random Tensors}
We also performed extensive benchmarks using networks composed of random tensors. To construct a random tensor, we draw each entry uniformly from the interval $[-1+\beta,\, 1+\beta]$, where $\beta$ is a positive bias parameter.

Setting $\beta = 0$ yields tensors with an even distribution of positive and negative entries; tensor networks in this regime are known to be computationally hard to contract\cite{Chen2025Sign} and typically do not converge to a BP fixed point under message passing. At the opposite extreme, setting $\beta = 1$ yields tensors whose entries are strictly positive; such networks are almost trivially easy to approximate, with the BP fixed point providing an almost exact estimate of the contraction.

Unless otherwise stated, we use a small positive bias of $\beta = 0.2$ when constructing the random tensors used in our examples. This choice produces networks for which BP typically converges to a fixed point, but that are still difficult enough such that the BP approximation exhibits substantial error (often between $1$--$10\%$), leaving sufficient room for improvement via the partitioned expansions.

\section{Message Symmetrization} \label{sect:sym}
Let us assume that a BP fixed point has been found for some tensor network $\mathcal T$, and let $\ket{\ovr \mu_i}$ and $\ket{\ovl \mu_i}$ denote the incoming and outgoing messages associated with index $i$ of the network. In order to form a suitable rank $r=1$ projector $P$ on index $i$, one must first make an appropriate change of gauge to \emph{symmetrize} these messages so that, after the gauge transformation, both the incoming and outgoing messages become equal to a single message $\ket{\mu_i}$. In this appendix we describe a general approach to achieve this symmetrization.

We begin by assuming that the incoming and outgoing messages have been normalized such that $\braket{\ovr \mu_i}{\ovl \mu_i} = 1$. We now consider a gauge transformation enacted by a pair of matrices $x^{-1}$ and $x$, producing new messages $\ket{\ovr \nu_i}$ and $\ket{\ovl \nu_i}$ defined as
\begin{align}
\bra{\ovr \nu_i} &\equiv \bra{\ovr \mu_i}\, x^{-1},\nonumber \\
\ket{\ovl \nu_i} &\equiv x\, \ket{\ovl \mu_i}. \label{eq:app1}
\end{align}
Our goal is to find a matrix $x$ such that these new messages coincide, $\ket{\ovr \nu_i} = \ket{\ovl \nu_i}$. For convenience, we choose the gauge such that the symmetrized message is the basis vector, $x \ket{\ovl \mu_i} = \ket{e_0}$ with $\ket{e_0} = [1, 0, 0, \ldots]^\dagger$.

Let index $i$ have dimension $d$. To construct $x$, we first identify a $(d-1)\times d$ matrix $R_\perp$ whose rows span the subspace orthogonal to $\ket{\ovl \mu_i}$. This matrix must satisfy
\begin{equation}
R_\perp \ket{\ovl \mu_i} = \tilde 0,
\end{equation}
with $\tilde 0$ the null vector, and must have orthonormal rows,
\begin{equation}
R_\perp R_\perp^\dagger = I_{d-1},
\end{equation}
where $I_{d-1}$ is the $(d-1)$-dimensional identity.

Such an $R_\perp$ can be constructed easily using an SVD. For example, in Python one may take the NumPy SVD of the row vector $\bra{\ovl \mu_i}$ using the option \verb|full_matrices=True|; the orthogonal complement $R_\perp$ is then built from the $(d-1)$ subleading rows of the returned \verb|Vh| matrix. We now construct the $(d\times d)$ gauge transformation matrix $x$ by stacking $\bra{\ovr \mu_i}$ on top of $R_\perp$:
\begin{equation}
x =
\begin{bmatrix}
\bra{\ovr \mu_i} \\
R_\perp
\end{bmatrix}.
\end{equation}
By construction, this matrix satisfies
\begin{equation}
x \ket{\ovl \mu_i} = \ket{e_0}.
\end{equation}
Moreover, since the first row of $x$ is $\bra{\ovr \mu_i}$, we also have
\begin{equation}
\bra{e_0}\, x = \bra{\ovr \mu_i}.
\end{equation}
Thus, using Eq.~\ref{eq:app1}, the re-gauged messages are symmetrized as desired:
\begin{equation}
\ket{\ovr \nu_i} = \ket{\ovl \nu_i} = \ket{e_0}.
\end{equation}

\section{Weight Passing} \label{sect:weight}
Many existing tensor network algorithms rely on constructing, for each index $i$, a diagonal matrix of positive weights $S_i$ that characterizes the relative importance of different index values. Examples include the simple-update PEPS algorithm\cite{Jiang2008Simple} and its generalization to arbitrary geometries\cite{Jahromi2019}, as well as certain implementations of the tensor renormalization group (TRG)\cite{Zhao2010RTN}. In these settings the weights $S_i$ provide an approximation to the local environment and can significantly improve the fidelity of local truncations. In the context of partitioned expansions, these weights $S_i$ can be used to construct rank-$r$ projectors $P$ by selecting the subspace associated with the largest $r$ weights. This offers an alternative to projectors derived from a BP fixed point, which are restricted to rank $r=1$. The iterative procedure described below, which we call \emph{weight passing}, provides a general method for constructing such weights on all indices of a tensor network. Our approach is similar to methods used in simple-update PEPS\cite{Jahromi2019} and some TRG variants\cite{Zhao2010RTN}, and is closely related to general tensor-network gauge-fixing techniques\cite{Evenbly2018Gauge}. It is also connected to re-gauging methods explored in the context of BP\cite{AlkabetzArad2021,Tindall2023Scipost}.

\begin{figure} [!t] 
\begin{center}
\includegraphics[width=7.0cm]{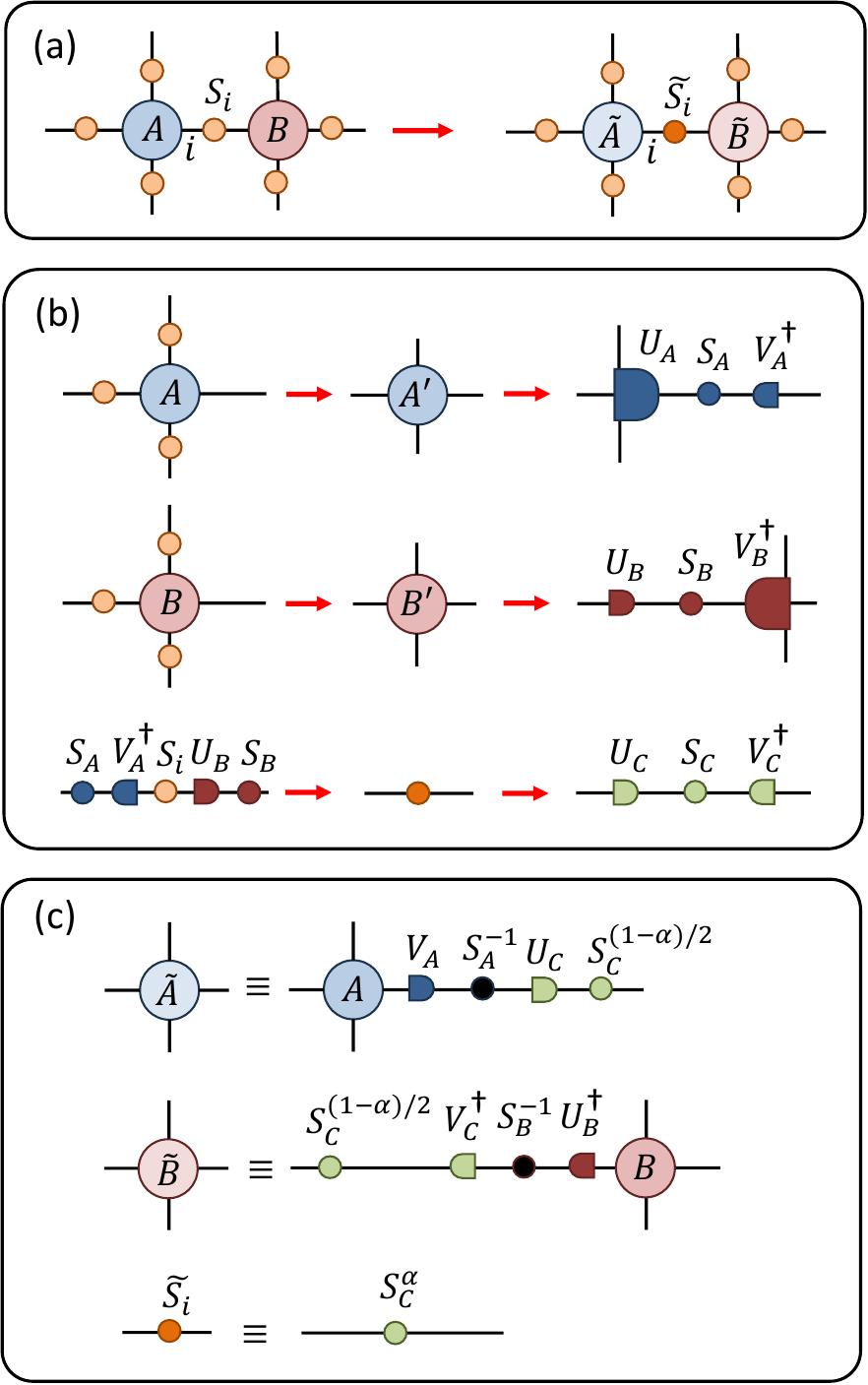}
\caption{(a) Given two tensors $A$ and $B$ sharing an index $i$ with weights $S_i$, an iteration of weight passing takes these to new tensors $\tilde A$, $\tilde B$ and $\tilde S_i$. (b) After absorbing environment weights, the singular value decomposition of tensors $A'$ and $B'$ is taken. Singular values $S_A$ and $S_B$ are combined with the existing index weights $S_i$, and a further SVD is taken. (c) Definition of new tensors $\tilde A$, $\tilde B$ and $\tilde S_i$.}
\label{fig:13}
\end{center}
\end{figure}

Suppose we are given a closed tensor network $\mathcal T$ that contracts to a scalar $\mathcal Z_0$. We begin by placing on each index $i$ a diagonal weight matrix $S_i$, initially set to the identity $I$. Consider a local patch of the network consisting of an index $i$ with its weight $S_i$, the two tensors $A$ and $B$ connected by that index, and the weights adjacent to those tensors, as depicted in Fig.~\ref{fig:13}(a). Our goal is to update this patch to new tensors $\{\tilde A, \tilde B, \tilde S_i\}$, leaving the remainder of the network unchanged, in such a way that the updated weight $\tilde S_i$ better reflects the contribution of each index value to $\mathcal Z_0$.

We first absorb the weights on all indices adjacent to $A$, except for index $i$, into tensor $A$ to obtain a modified tensor $A'$. Likewise, we absorb the weights adjacent to $B$ into tensor $B$ to obtain $B'$. Each of these tensors is then decomposed via an SVD (treating each tensor as a matrix between index $i$ and all other indices),
\begin{align}
A' &\rightarrow U_A S_A V_A^\dagger , \nonumber\\
B' &\rightarrow U_B S_B V_B^\dagger , \label{eq:weight1}
\end{align}
as illustrated in Fig.~\ref{fig:13}(b). The singular values from Eq.~\ref{eq:weight1} are then contracted with the current weight $S_i$, and a second SVD is performed,
\begin{equation}
S_A V_A^\dagger S_i U_B S_B \;\rightarrow\; U_C S_C V_C^\dagger, \label{eq:weight2}
\end{equation}
see Fig.~\ref{fig:13}(b).

We now specify how to define the updated weight $\tilde S_i$. One option is to take $\tilde S_i = S_C$ from Eq.~\ref{eq:weight2}, but this typically leads to divergence when the update is iterated. Instead we take a power $\alpha$ of the weights, defining $\tilde S_i = S_C^\alpha$ with $\alpha \in [0,1]$. The parameter $\alpha$ controls the sharpness of the weight distribution: for $\alpha=0$ the weights remain completely flat, $\tilde S_i = I$, while $\alpha \rightarrow 1$ increasingly sharpens the weight spectrum. The updated tensors $\tilde A$ and $\tilde B$ are chosen so that the transformation constitutes a pure gauge change, ensuring the contracted value of the network remains unchanged:
\begin{align}
\tilde A &= A \, V_A \, S_A^{-1} \, U_C \, S_C^{(1-\alpha)/2}, \nonumber\\
\tilde B &= S_C^{(1-\alpha)/2} \, V_C^\dagger \, S_B^{-1} \, U_B^\dagger \, B, \nonumber\\
\tilde S_i &= S_C^\alpha , \label{eq:weight3}
\end{align}
as depicted in Fig.~\ref{fig:13}(c). Notice that, when $\alpha < 1$, part of the information contained in $S_C$ is distributed into the neighboring tensors via the factors $S_C^{(1-\alpha)/2}$. 

The weight passing algorithm consists of performing the update in Eq.~\ref{eq:weight3} for each index $i$ in a sweep over the network, and iterating such sweeps until all weights converge. The benchmark results presented in Sect.~\ref{sect:beyond}, which employed higher-rank projectors, were obtained using the weight passing algorithm with $\alpha = 0.8$. For these benchmark tests the method converged reliably, requiring a number of iterations comparable to that needed for BP to reach a fixed point. We remark that when the algorithm is run with $\alpha = 1$, the weights on each index typically become increasingly peaked as the iterations proceed, with sub-leading values approaching zero, eventually leading to numerical instability due to floating-point precision limits.

We also note several empirical observations. First, during testing, we observed some networks for which weight passing converged while BP message passing did not, with BP messages continuing to fluctuate strongly even after many iterations. Examples included networks of random tensors with only a very small positive bias on the tensor elements. Second, convergence of the weight passing could often be improved by first running the algorithm with a smaller value of $\alpha$, producing a flatter weight spectrum, and then re-running with larger $\alpha$. We speculate that this “ramping-up’’ of $\alpha$ mimics annealing, gradually steering the system toward a “low-energy’’ configuration. Finally, the weight passing algorithm was also observed to fail for certain difficult networks, such as networks of random tensors without any positive bias, for which the weights remained nearly flat even after many iterations. In such cases, expansions based on these weights yield poor accuracy. A more comprehensive comparison between BP message passing and the weight passing algorithm is beyond the scope of this manuscript and will be explored in future work.

\bibliographystyle{apsrev4-2}
\bibliography{main}

\end{document}